\documentclass[
 reprint,
 superscriptaddress,
nofootinbib,
 amsmath,amssymb,
]{revtex4-2}

\usepackage{graphicx}
\usepackage{dcolumn}
\usepackage{bm}

\usepackage{longtable,color,array}
\usepackage[unicode=true,pdfusetitle,
bookmarks=true,bookmarksnumbered=true,bookmarksopen=true,bookmarksopenlevel=1,
breaklinks=false,pdfborder={0 0 0},backref=false,colorlinks=true]{hyperref}
\hypersetup{citecolor=blue,filecolor=blue,linkcolor=blue,urlcolor=blue,pdfauthor={Name}}
\usepackage[usenames,dvipsnames]{xcolor}
\usepackage{latexsym,amsmath,amssymb,amsfonts,mathrsfs}
\usepackage{epsfig,subfigure,placeins,float}
\usepackage{booktabs,longtable,multirow}
\usepackage{exscale,relsize}
\usepackage[normalem]{ulem}
\usepackage[T1]{fontenc}
\usepackage[utf8]{inputenc}
\usepackage{times}
\usepackage{tikz}
\usepackage{aas_macros}
\usepackage{tabularx}
\usepackage{tabularray}
\usepackage{diagbox}
\newcolumntype{M}[1]{>{\centering\arraybackslash}m{#1}}
\usepackage{blkarray}

\allowdisplaybreaks
\usepackage{colortbl}
\usepackage{pifont}
    \newcommand{\cmark}{\ding{51}}
    \newcommand{\xmark}{\ding{55}}
\usepackage{xfrac}
\usepackage[noabbrev]{cleveref}
\usepackage{fontawesome5}

\newcommand{\comment}[1]{}
\newcommand{\red}[1]{#1}
\newcommand{\blue}[1]{#1}

\begin{document}


\title{
A Master Equation for Screening in Luminal Horndeski Gravity
}

\author{Sergi Sirera}
 \email{sergi.sirera@port.ac.uk}
 \affiliation{Institute of Cosmology \& Gravitation, University of Portsmouth, Portsmouth, PO1 3FX, U.K.}

\author{Tessa Baker}
 \affiliation{Institute of Cosmology \& Gravitation, University of Portsmouth, Portsmouth, PO1 3FX, U.K.}

\author{James Hallam}
 \affiliation{Institute of Cosmology \& Gravitation, University of Portsmouth, Portsmouth, PO1 3FX, U.K.}

\author{Krishna Naidoo}
 \affiliation{Institute of Cosmology \& Gravitation, University of Portsmouth, Portsmouth, PO1 3FX, U.K.}
 \affiliation{Max-Planck-Institut f\"ur Astronomie, K\"onigstuhl 17, 69117 Heidelberg, Germany}

\date{\today}

\begin{abstract}
Determining the active screening mechanism from a general scalar-tensor Lagrangian remains a challenging problem.
As a diagnostic tool, we present a systematic study of nonlinear cosmological perturbations in luminal Horndeski theories. Working in the $\alpha$-basis on a flat FLRW background, we derive and organise the full set of unapproximated second-order perturbation equations, and systematically apply the quasi-static and weak-field limits.
We find that second-order effects modify only the scalar field equation.
We derive, for static and spherically symmetric configurations, a master screening equation \red{that classifies the nonlinear operators driving screening, recovering the Vainshtein mechanism and the onset of Chameleon screening.} We also identify a novel \red{candidate} regime, which we term Phaedrus screening, characterised by a screening radius that scales linearly with the source mass. For each mechanism, we derive analytical and numerical solutions and clarify the conditions under which they activate.
Two new publicly available software packages are introduced: i) \href{https://github.com/sergisl/xAlpha}{\faGithub \;xAlpha}, a \texttt{Mathematica} package to compute and organise perturbation equations in scalar-tensor theories, and ii) \href{https://github.com/sergisl/escut}{\faGithub \;escut}, a \texttt{Python} module to solve the nonlinear scalar equation.
In many cases, these tools enable the identification of the active screening type directly from a luminal Horndeski Lagrangian.
\end{abstract}

\maketitle

\tableofcontents
\nocite{DESI:2024mwx, DES:2025sig, Koyama:2015vza, Ferreira:2019xrr, 
        DESI:2025zgx, DESI:2025fii, DESI:2025ejh, Naidoo:2026umv, 
        Hallam:2026qsk, Ye:2024ywg, Wolf:2024stt, Wolf:2024eph, 
        Gu:2025xie, 1993tegp.book.....W, Bertotti:2003rm, Will:2014kxa, 
        Burrage:2020jkj, Brax:2013ida, Joyce:2014kja, Babichev:2013usa, 
        Khoury:2003aq, Hinterbichler:2010es, Babichev:2009ee, Vainshtein:1972sx, 
        Hu:2007nk, PhysRevD.80.123003, Barreira:2013eea, Brax:2014yla, 
        Jain:2012tn, laureijs2011euclid, LSST:2008ijt, Wolf:2023uno, 
        Ferreira:2025fpn, DeFelice:2010as, DeFelice:2011hq, Gubitosi:2012hu, 
        Bellini:2014fua, Zumalacarregui:2016pph, Baker:2019gxo}
\section{Introduction}\label{sec-intro}
\begin{table*}
  \begin{tabular*}{0.7\textwidth}{@{\extracolsep{\fill}} lcc}
    \toprule
    Theory & Nonlinear & No approximations \\
    \midrule
    Horndeski \cite{DeFelice:2011hq,Bellini:2014fua}, Beyond Horndeski \cite{Gleyzes:2014rba,DAmico:2016ntq} & \xmark\ ($1^{\rm st}$) & \cmark\\
    Luminal Horndeski \cite{Amendola:2025xka}, Horndeski \cite{Kimura:2011dc,Cusin:2017mzw}
    & \cmark\ ($3^{\rm rd}$) & \xmark\\
    Luminal Horndeski (this work) & \cmark\ ($2^{\rm nd}$) & \cmark\\
    \bottomrule
  \end{tabular*}
  \caption{Summary review of analytical cosmological perturbation studies of general ST theories. Approximations correspond to i) the quasistatic approximation (QSA), ii) the weak-field limit, and iii) the assumption of Vainshtein providing the dominant screening.
  Later on, after showing the full unapproximated expressions, we also impose weak-field and QSA limits.
  In this work, we use the $\alpha$-basis \cite{Bellini:2014fua}, also adopted in \cite{Gleyzes:2014rba,DAmico:2016ntq,Cusin:2017mzw}, while other works employ alternative parameterisations \cite{Arjona:2019rfn}. Note that this Table is restricted to general ST frameworks; other analytical derivations of nonlinear perturbations exist for specific models such as DGP \cite{Koyama:2007ih}, $f(R)$ \cite{Koyama_2009} and Galileons \cite{Bartolo:2013ws}.
  }
     \label{lit-rev}
\end{table*}
The unresolved nature of dark energy and persistent tensions between various observational probes and the standard cosmological model ($\Lambda$CDM) \cite{DESI:2024mwx,DES:2025sig} collectively provide a strong motivation for investigating modified theories of gravity \cite{Koyama:2015vza,Ferreira:2019xrr}.  
Among these, scalar-tensor (ST) theories stand out as particularly promising candidates. By introducing a dynamical scalar field, these theories can impact both the expansion history and the growth of large-scale structure (LSS). Hence, they offer a rich framework to test the fundamental laws of gravity and potentially alleviate the aforementioned cosmological tensions.
In particular, recent data \cite{DESI:2024mwx,DESI:2025zgx,DESI:2025fii,DESI:2025ejh} modestly favours models featuring phantom-crossing behaviour over standard $\Lambda$CDM and other minimal ST alternatives.\footnote{See for instance the so-called asymptotic cubic Galileon models \cite{Naidoo:2026umv,Hallam:2026qsk} and non-minimally coupled theories \cite{Ye:2024ywg,Wolf:2024stt,Wolf:2024eph,Gu:2025xie}.}
Crucially, for these theories to remain viable, they must recover the well-tested and highly precise predictions of General Relativity (GR) on Solar System scales, see e.g. \cite{1993tegp.book.....W,Bertotti:2003rm,Will:2014kxa,Burrage:2020jkj}. Thus, non-trivial mechanisms \textit{screening} the effect of the scalar on nonlinear (shorter, denser) scales must be naturally activated \cite{Brax:2013ida,Joyce:2014kja,Babichev:2013usa}.

Several qualitatively distinct screening mechanisms have been found, including the Chameleon \cite{Khoury:2003aq}, Symmetron \cite{Hinterbichler:2010es}, $K$-mouflage \cite{Babichev:2009ee}, and Vainshtein \cite{Vainshtein:1972sx}, each relying on different nonlinear operators to suppress scalar interactions in dense or high-curvature environments. While their operation has been extensively studied in specific models, see e.g. \cite{Hu:2007nk,PhysRevD.80.123003,Barreira:2013eea,Brax:2014yla,Jain:2012tn}, a unified and systematic identification of the operators responsible for screening has not yet been fully established. In other words, given a general ST Lagrangian, it remains challenging to ascertain which screening mechanism is active, and to what extent.

A unified screening framework is more than a mathematical nicety; it is a prerequisite for the robust testing of gravity with Stage IV LSS surveys such as Euclid \cite{laureijs2011euclid} and LSST \cite{LSST:2008ijt}. Because many viable modified gravity models predict identical cosmic expansion histories at the background and linear levels \cite{Wolf:2023uno,Ferreira:2025fpn}, they remain otherwise indistinguishable.\footnote{This degeneracy is termed `permanent underdetermination' (see \cite{Wolf:2025dss} for a taxonomy of potential resolutions). Alongside complementary tests in strong gravity regimes, as is possible e.g. in the context of black hole quasinormal modes in \cite{Smulders:2026bya}, the screening identification framework presented in this work can provide a way to discriminate between otherwise identical theories.} Consequently, as these surveys achieve their promised precision, breaking these model degeneracies requires moving beyond well-understood linear perturbation theory \cite{DeFelice:2010as,DeFelice:2011hq,Gubitosi:2012hu,Bellini:2014fua,Zumalacarregui:2016pph} directly into the nonlinear regime. By capturing the transition where the scalar field is suppressed, one can isolate the characteristic quasi-nonlinear imprints on matter clustering and the lensing potential \cite{Baker:2019gxo}, establishing screening as a primary observable to distinguish between fundamental gravity theories.

Within generalised ST theories, previous analytical work on cosmological perturbations has focused either on linear treatments \cite{DeFelice:2011hq,Bellini:2014fua}, or on nonlinear extensions under strong approximations \cite{Kimura:2011dc,Cusin:2017mzw}, see Table~\ref{lit-rev}. In this work, we fill this gap by providing the full, organised second-order structure of perturbation equations for luminal Horndeski theories without imposing such approximations.\footnote{Note that we consider exclusively scalar perturbations of the metric, assuming vector and tensor modes to be negligible. Vector perturbations decay with the expansion of the Universe, while tensor modes are subdominant in the late-time dynamics of structure formation.
At nonlinear order, the decomposition breaks down. While relevant for other contexts, see e.g. \cite{Ananda:2006af,Baumann:2007zm,Creminelli:2018xsv,Creminelli:2019kjy,Bari:2023rcw}, these mixed couplings remain sub-leading for the screening dynamics investigated here.
}
Here, we truncate the perturbative expansion at second order, which has specific implications for each mechanism: first, it fully captures Vainshtein screening within the luminal subclass, as this is driven exclusively by the cubic Galileon and thus fully described by quadratic nonlinearities.
Second, it also provides an effective description of Chameleon screening by capturing the leading-order shift in the scalar's effective mass and density-dependent shift of the potential minimum. While formally resolving the exact thin-shell profile deep inside dense regions requires an infinite series (because the local perturbation ultimately grows comparable to the background) our truncated framework successfully isolates the physical onset of the mechanism.
However, mechanisms relying on higher-order nonlinearities, e.g. $K$-mouflage (from $(\partial \phi)^4$) and the Symmetron (from a $\phi^4$ potential), manifest primarily at third order. We have verified this for $K$-mouflage via preliminary third-order computations in a restricted ($K$ and $G_4$) subclass, but leave the formal organisation of the full third-order system for future work.\footnote{\red{Although the full Symmetron screening profile requires third order, the density-dependent symmetry restoration that triggers the mechanism is already captured at second order, via the nonlinear coupling $\delta \phi \nabla^2 \Phi$.}}
\red{For the candidate Phaedrus a caveat applies: the same non-canonical kinetic functions that generate it can also source higher-order ($K$-mouflage-type) interactions. Establishing that the quadratic terms can dominate is therefore a non-trivial requirement which we explicitly examine and identify in Sec.~\ref{sec:phaedrus}.}

Finally, let us highlight that the computational cost of full N-body simulations makes them an impractical option for exploring the vast parameter space of beyond-$\Lambda$CDM scenarios. Instead, fast hybrid simulation techniques (such as \texttt{Hi-COLA} \cite{Wright:2022krq,Gupta:2024seu} and \texttt{PySCo-EFT} \cite{Ganjoo:2026ugf}) and semi-analytical models (such as \texttt{ReAct} \cite{Cataneo:2018cic,Bose:2020wch,Bose:2022vwi}) offer a viable and efficient alternative for the analysis of Stage IV survey data. However, these tools rely on analytical inputs to model nonlinear interactions. The perturbative framework developed in this work (see also e.g. \cite{Aviles:2018qot}) is designed to eventually provide precisely this required theoretical foundation.

\textbf{Outline:}
The remainder of this paper is organised as follows. Section~\ref{sec:luminalHorndeski} provides a brief review of luminal Horndeski gravity at the background level. In Section~\ref{sec:nonlinear}, we present the general structure of the second-order perturbation equations. We demonstrate how these reduce to the effective modified Poisson, gravitational slip, and scalar field equations, ultimately using them \red{to organise the second-order nonlinear operators into a single unified screening equation.} Section~\ref{sec:screening} applies this master equation to a spherically symmetric, static source, showing that it successfully recovers both the Vainshtein and Chameleon mechanisms. Furthermore, we identify \red{a candidate new regime, Phaedrus screening, detailing its unique features and the open questions bearing on its physical viability.} Finally, we conclude in Section~\ref{sec:conclusions} with a summary of our findings and an outlook on future directions.

\section{Luminal Scalar-Tensor Gravity}\label{sec:luminalHorndeski}
Horndeski gravity is the most general ST theory manifestly leading to second-order equations of motion~\cite{Horndeski:1974wa,Deffayet:2011gz,Kobayashi:2011nu}.\footnote{Extensions exist, namely beyond-Horndeski~\cite{Gleyzes:2014dya} and DHOST (Degenerate Higher-Order Scalar-Tensor) theories~\cite{Langlois:2015cwa,Crisostomi:2016czh,BenAchour:2016fzp}, which involve higher-order equations of motion but nonetheless remain ghost-free due to a degeneracy in the Lagrangian.}
It provides a general consistent construction to explore modifications of GR, motivated in part by its ability to provide dynamical cosmic acceleration as well as evading Solar System constraints.
More broadly, it constitutes a framework to generically test gravitational interactions, see e.g. \cite{Kobayashi:2019hrl} for a review.

Among the many predictions of such generalised ST theories, one is that gravitational waves (GWs) may travel at speeds different than light, with deviations typically quantified by the parameter $\alpha_T$. This quantity has been strongly constrained by the GW170817/GRB 170817A event to $|\alpha_T|\leq10^{-15}$ \cite{LIGOScientific:2017zic,Goldstein:2017mmi,Savchenko_2017}, which has been taken to imply that Horndeski theories require $\alpha_T=0$ to remain viable\footnote{
Nonetheless, it is important to stress that this observation occurred at a frequency roughly matching the cutoff scale of such theories when interpreted as dark energy models (i.e. $f\sim\Lambda_3\sim\mathcal{O}(10^2)$Hz) \cite{deRham:2018red}. Hence, it remains not entirely straightforward to impose such constraint on a cosmological setting, see related discussions in~\cite{Harry:2022zey,Baker:2022rhh,Baker:2022eiz,Sirera:2023pbs,Atkins:2024nvl}. Here, we take the conservative approach and understand this observation to seriously imply $\alpha_T=0$, i.e. obeying $c_{GW}=c$.
}
\cite{Creminelli:2017sry,Ezquiaga:2017ekz,Baker:2017hug}, resulting in the luminal Horndeski action
\begin{align}
S=\int d^4x\sqrt{-g}\Big[&G_4(\phi)R+K(\phi, X) \nonumber\\
&\qquad-G_3(\phi, X)\Box\phi+{\cal L}_{\rm m}\Big],\label{action}
\end{align}
where $\mathcal{L}_{\rm m}$ is the matter Lagrangian.
Above, we have introduced the shorthand $X\equiv-\frac{1}{2}\phi_\mu\phi^\mu$ for the kinetic term of the scalar, where $\phi_\mu\equiv\nabla_\mu\phi$ (and $\phi_{\mu\nu}\equiv\nabla_\nu\nabla_\mu\phi$), and $K$ and $G_i$ are free functions of $\phi$ and $X$, with $G_{iX}$ denoting the partial derivative of $G_i$ with respect to $X$. For the theory to be cosmologically relevant, the scales of each operator are set to yield $\mathcal{O}(1)$ contributions to the cosmological background evolution (i.e. to the Friedmann equations).\footnote{Specifically, we implicitly assume standard cosmological mass-scale matching: $K \sim \Lambda_2^4 \tilde{K}$, $G_3 \sim (\Lambda_2^4/\Lambda_3^3)\tilde{G}_3$, and $G_4 \sim M_P^2 \tilde{G}_4$, where the dimensionless tilded functions are $\mathcal{O}(1)$ and $\Lambda_2^2 \equiv M_P H_0$ and $\Lambda_3^3 \equiv M_P H_0^2$. Note that \( M_P \gg \Lambda_2 \gg \Lambda_3 \), and thus the lowest of these, \( \Lambda_3 \), sets the cutoff scale of the theory.}

Finally, we assume that matter is minimally coupled to the spacetime metric and described by a perfect fluid with a stress-energy tensor of the form
\begin{align}
    T_{\mu\nu} &\equiv -\frac{2}{\sqrt{-g}} \frac{\delta (\sqrt{-g} {\cal L}_{\rm m})}{\delta g^{\mu\nu}}= (\rho + p) u_\mu u_\nu + p g_{\mu\nu}.
\end{align}
Here, $\rho$ is the energy density, $p$ is the isotropic pressure, and $u_\mu$ is the (non-relativistic) four-velocity of the fluid.

\subsection{Effective functions for cosmological perturbations}
A useful parameterisation of linear perturbations in ST theories is provided by the $\alpha$ functions, which capture the time evolution of background quantities and their impact on cosmological dynamics \cite{Bellini:2014fua}. For the theory described in \eqref{action}, these take the form\footnote{As mentioned, one additionally has $\alpha_T$ (tensor speed excess) for full Horndeski theories, and $\alpha_H$ for Beyond Horndeski theories \cite{Gleyzes:2014rba}.}
\begin{align}
    \alpha_M&=\frac{1}{HM_*^2}\frac{dM_*^2}{dt}=\frac{2\dot\phi G_{4\phi}}{HM_*^2},\\
    \alpha_K&=\frac{2X}{H^2M_*^2}[K_X+2XK_{XX}-2G_{3\phi}-2XG_{3\phi X}\nonumber\\
    &\qquad\qquad\qquad+6\dot\phi H(G_{3X}+XG_{3XX})],\\
    \alpha_B&=\frac{2\dot\phi}{HM_*^2}(XG_{3X}-G_{4\phi}),
\end{align}
where we have used $M^2_*= 2 G_4$.
These three functions, respectively referred to as the \textit{Planck-mass running}, the \textit{kineticity}, and the \textit{braiding}, each characterise a distinct physical aspect of the underlying ST dynamics \cite{Bellini:2014fua}.

In this work, we present the second-order perturbation equations expressed in terms of the $\alpha$ functions. We demonstrate that, for luminal Horndeski theories, the standard set $\{\alpha_K, \alpha_B, \alpha_M\}$ and their derivatives are sufficient to fully characterise second-order perturbations.\footnote{Note that one specific term, related to the nonlinear contribution to the effective mass of the scalar perturbation, requires special treatment and is therefore not fully converted to $\alpha$ functions. More details can be found in Appendix~\ref{app:expressions}.}
This is not guaranteed in general; as found in \cite{Cusin:2017mzw}, non-luminal ST models require the definition of new functions beyond the linear set to fully characterise nonlinear perturbation equations. Despite this sufficiency, the resulting expressions here grow in complexity. To simplify the notation, we therefore introduce the following set of $\gamma$ functions, which absorb specific combinations of the $\alpha$ parameters:
\begin{align}
    \{\gamma_E,\gamma_M,\gamma_K,\gamma_B,\gamma_A,\gamma_F,\gamma_D,\gamma_C,\gamma_X\},
\end{align}
with the full expressions given in Appendix~\ref{app:expressions}.

\subsection{Cosmological background evolution}
We assume a flat Friedmann-Lemaître-Robertson-Walker (FLRW) metric and a time-dependent scalar field, given by
\begin{align}
    ds^2&=-dt^2+a^2(t)d\mathbf{x}^2, & \phi&=\phi(t),
    \label{eq:bg}
\end{align}
where $a(t)$ is the scale factor, and $H(t)=\dot{a}/a$ is the Hubble parameter.\footnote{Note that we have used the definition $\dot{a}\equiv\frac{da}{dt}$.}

The metric and scalar equations of motion for this theory are shown in covariant form in Appendix~\ref{app:cov-eoms}. When imposing the cosmological background above, these inherit the $1+3$ splitting of spacetime coordinates, and are given by
\begin{align}
\begin{bmatrix}
\mathcal{E} &
0 \\[6pt]
0 &
a^2\delta_{ij}\mathcal{P}
\end{bmatrix}
=
\begin{bmatrix}
-\rho_m & 0 \\[6pt]
0 & \red{-a^2\delta_{ij}\,}p_m
\end{bmatrix},
& &\mathcal{S}=0,
\label{eq:bg-eqs}
\end{align}
\red{where $\delta_{ij}$ is the Kronecker delta, the background is homogeneous, and so the spatial equation is isotropic.}
Above, we have defined the following terms (agreeing with~\cite{Kimura:2011dc})\footnote{We identify a sign discrepancy in the $G_{3\phi X}$ term of Eq.~\eqref{eq:scalar-backg} relative to Eq.~(A.5) in \cite{Bellini:2014fua}. This mismatch is similarly evident when comparing Eq.~(9) of \cite{Kimura:2011dc} to the results in \cite{Bellini:2014fua}. \red{The full step-by-step derivation is available in the accompanying \href{https://github.com/sergisl/xAlpha}{\faGithub \;xAlpha} notebook. Throughout our derivations, we employ the background identity $2X=\dot \phi^2$. We also note the relation $X\mathcal{E}_X=2H^2G_4^2(\alpha_K+3\alpha_B)$.}}
\begin{align}
    {\cal E}&\equiv 2X(K_X-G_{3\phi}+3H\dot{\phi}G_{3X})-K\nonumber\\
    &\quad-6H(HG_4+\dot\phi G_{4\phi })\\
    {\cal P}&\equiv K-2X\left(G_{3\phi}+\ddot\phi G_{3X} -2G_{4\phi\phi}\right)
    \nonumber\\
    &\quad+2\left(3H^2+2\dot H\right) G_4+2\left(\ddot\phi+2H\dot\phi\right) G_{4\phi},\\
    \mathcal{S}&\equiv\ddot{\phi} \Big[K_X-2G_{3\phi}+6H\dot{\phi}\left(G_{3X}+X G_{3XX}\right)\nonumber\\
    &\quad+2X \left(K_{XX}- G_{3\phi X}\right)\Big]-K_{\phi}-6 G_{4\phi}\left(\dot{H}+2 H^2\right)\nonumber\\
    &\quad+3H\dot{\phi}\left(K_{X}-2 G_{3\phi}+2 X G_{3\phi X}\right)\nonumber\\
    &\quad+2X\left[K_{\phi X}-G_{3\phi\phi}+3(3H^2+\dot H)G_{3X}\right].
    \label{eq:scalar-backg}
\end{align}

For convenience, we will normalise the following quantities by $M_*^2$ and redefine them as
\begin{align}
    \tilde{\mathcal{E}}&=\frac{\mathcal{E}}{M_*^2}, & \tilde{\mathcal{P}}&=\frac{\mathcal{P}}{M_*^2}, & \tilde{\mathcal{S}}&=\frac{\mathcal{S}}{M_*^2},\nonumber\\ \tilde{\rho}_m&=\frac{\rho_m}{M_*^2}, &
    \tilde{p}_m&=\frac{p_m}{M_*^2}
    \label{eq:tilded-expressions}
\end{align}
Finally, the conservation equation $\nabla_\mu T^{\mu\nu} = 0$ results in\footnote{Equivalently, we have
\begin{align}
    \dot{\tilde{\mathcal{E}}} + 3H(\tilde{\mathcal{E}} + \tilde{\mathcal{P}}) = \alpha_M H \tilde{\rho}_m + \dot{\phi} \tilde{\mathcal{S}}.
\end{align}
}
\begin{align}
    \dot{\mathcal{E}} + 3H(\mathcal{E} + \mathcal{P}) = \dot{\phi} \mathcal{S}.
\end{align}
In order for this background solution to be a physically meaningful description, stability conditions need to be satisfied. These are collected for completeness in Appendix~\ref{app:cov-eoms}.

\section{Nonlinear perturbation equations}\label{sec:nonlinear}
Let us now consider perturbations on top of this background. Working in the Newtonian gauge \cite{1992PhR...215..203M,Ma:1995ey}, the perturbed metric is written as
\begin{align}
ds^2=-(1+2\Phi)dt^2+a^2(1-2\Psi)d\mathbf{x}^2,
\end{align}
where $\Phi$ and $\Psi$ represent the scalar metric potentials. The corresponding perturbations in the scalar field and the matter sector are defined as
\begin{align}
\phi(t, \mathbf{x}) &= \phi(t) + \delta\phi(t, \mathbf{x}), \\
\rho_{\rm m}(t, \mathbf{x}) &= \rho_{\rm m}(t) [1 + \delta(t, \mathbf{x})], \\
p_{\rm m}(t, \mathbf{x}) &= p_{\rm m}(t) + \delta p_{\rm m}(t, \mathbf{x}),
\end{align}
where $\phi(t)$, $\rho_{\rm m}(t)$, and $p_{\rm m}(t)$ denote the background values.
In what follows, we use the dimensionless quantity
\begin{align}
Q\equiv H\frac{\delta\phi}{\dot\phi}=\frac{\delta\phi}{d\phi/d\ln a}.\label{eq:Qdef}
\end{align}
\vspace{0pt}

Several conventions exist in the literature for defining scalar field perturbations. In particular, the standard perturbation $\delta\phi$ is used in~\cite{DeFelice:2011hq}, while the velocity potential $v_X=-\delta\phi/\dot\phi$ is used in \cite{Bellini:2014fua}. Here, following  \cite{Kimura:2011dc,Wright:2022krq} we use the dimensionless $Q$ as defined above. Importantly, such choice has implications on the form of perturbation equations. For instance, $\ddot Q$ terms are partially converted into effective mass terms when expressed via $\delta\phi$ or $v_X$. We detail the explicit mappings between these conventions in Appendix~\ref{app:conversions}, though we emphasise that they ultimately yield equivalent predictions for physical observables~\cite{Pace:2020qpj}.

By introducing the book-keeping parameter $\epsilon$, \red{which counts the amplitude of the metric and scalar-field perturbations,} we can schematically write the metric EOMs to second order in these perturbations as
\begin{align}
&\begin{bmatrix}
\mathcal{E} + \epsilon \mathcal{E}^{(1)} + \epsilon^2 \mathcal{E}^{(2)} &
\epsilon \mathcal{A}^{(1)}_i + \epsilon^2 \mathcal{A}^{(2)}_i \\[6pt]
\epsilon \mathcal{A}^{(1)}_i + \epsilon^2 \mathcal{A}^{(2)}_i &
a^2\!\left(\delta_{ij}\mathcal{P} + \epsilon \mathcal{P}^{(1)}_{ij} + \epsilon^2 \mathcal{P}^{(2)}_{ij}\right)
\end{bmatrix}\nonumber \\
&\qquad =
\begin{bmatrix}
-\bar{\rho}_m(1 + \delta) & \delta T_{0i} \\[6pt]
\delta T_{i0} & \red{-a^2\delta_{ij}(}\bar{p}_m \red{+} \delta p_m\red{)}
\end{bmatrix},
\label{eq:matrix-eoms}
\end{align}
and similarly, the scalar EOM as
\begin{equation}
    \mathcal{S} + \epsilon\mathcal{S}^{(1)} + \epsilon^2\mathcal{S}^{(2)} = 0.
    \label{eq:scalar-eom}
\end{equation}
\red{Note that the expansion parameter $\epsilon$ counts only the amplitude of the metric and scalar-field perturbations, while the matter sector is instead retained as a full, non-linear source (e.g. the density contrast $\delta$ is not assumed to be small and reaches in fact $\delta\sim\mathcal{O}(10^2)$ on cluster scales).}

Having constructed the perturbation equations as such, we can now write, in full generality, each of the linear and quadratic terms above for the luminal Horndeski action \eqref{action}.
To do so, we collect the perturbation fields in the following vector
\begin{align}
Y^a = \begin{pmatrix} \Phi \\ \Psi \\ Q \end{pmatrix},
\end{align}
and define the derivative operator $\mathcal{D}_{ij}\equiv\delta_{ij}\nabla^2-\partial_i\partial_j$.
We can then write the full perturbation equations as
\begin{widetext}
    \begin{align}
    \tilde{\mathcal{E}}^{(1)}&=\sum_a \left(H^2A^a_1\cdot Y^a+HA^a_{2}\cdot\dot Y^a-\frac{1}{a^2}A^a_{3}\cdot \nabla^2Y^a\right),\label{eq:E1}\\
    \tilde{\mathcal{E}}^{(2)}&=\sum_{a,b}\bigg(\frac{H^2}{2}A^{(ab)}_1\cdot Y^aY^b+HA^{ab}_2\cdot \dot Y^a Y^b-3A^{(ab)}_{3}\cdot \dot Y^a\dot Y^b-\frac{2}{a^2}A^{ab}_{4}\cdot Y^a\nabla^2 Y^b-\frac{1}{2a^2}A^{(ab)}_{5}\cdot \partial_i Y^a\partial^i Y^b\nonumber\\
    &\qquad\qquad+\frac{1}{Ha^2}A^{ab}_{6}\cdot \dot Y^a\nabla^2 Y^b\bigg),\label{eq:E2}\\
    \tilde{\mathcal{A}}^{(1)}_i&=\sum_a\left(HB^a_{1}\cdot\partial_i Y^a-B^a_{2}\cdot\partial_i\dot Y^a\right),\label{eq:A1}\\
    \tilde{\mathcal{A}}^{(2)}_i&=\sum_{a,b}\Big(HB^{ab}_{1}\cdot Y^a\partial_i Y^b-B^{ab}_{2}\cdot\dot Y^a\partial_i Y^b-B^{ab}_{3}\cdot Y^a\partial_i\dot Y^b+\frac{1}{H}B^{ab}_{4}\cdot\dot Y^a\partial_i \dot Y^b-\frac{2}{Ha^2}B^{ab}_{5}\cdot\partial^j Y^a\mathcal{D}_{ij} Y^b\Big),\label{eq:A2}\\
    \tilde{\mathcal{P}}^{(1)}_{ij}&=\sum_a\Big[\delta_{ij}\left(H^2a^2C^a_1\cdot Y^a+Ha^2C^a_2\cdot \dot Y^a-a^2C^a_{3}\cdot\ddot Y^a\right)-C^a_{4}\cdot\mathcal{D}_{ij}Y^a\Big],\label{eq:P1}\\
    \tilde{\mathcal{P}}^{(2)}_{ij}&=\sum_{a,b}\Big[\delta_{ij}\Big(-H^2a^2C^{(ab)}_1\cdot Y^a Y^b+Ha^2C^{ab}_2\cdot Y^a\dot Y^b-a^2C^{(ab)}_{3}\cdot\dot Y^a\dot Y^b+a^2C^{ab}_{4}\cdot Y^a\ddot Y^b+\frac{a^2}{H}C^{ab}_{5}\cdot\dot Y^a\ddot Y^b\nonumber\\
    &\qquad\qquad+C^{(ab)}_{6}\cdot\partial_k Y^a\partial^k Y^b+\frac{2}{H}C^{ab}_{7}\cdot\partial_k Y^a\partial^k \dot Y^b\Big)+ C^{(ab)}_{8}\cdot\partial_i Y^a\partial_j Y^b\nonumber\\
    &\qquad\qquad-\frac{1}{H}C^{ab}_{9}\cdot\partial_i Y^a\partial_j\dot Y^b+C^{ab}_{10}\cdot Y^a\mathcal{D}_{ij}Y^b\Big],\label{eq:P2}\\
    \tilde{\mathcal{S}}^{(1)}&=\sum_a -\frac{H}{\dot\phi}\left(H^2D^a_1\cdot Y^a+HD^a_{2}\cdot\dot Y^a+D^a_{3}\cdot\ddot Y^a+\frac{1}{a^2}D^a_{4}\cdot \nabla^2Y^a\right),\label{eq:S1}\\
    \tilde{\mathcal{S}}^{(2)}&=\sum_{a,b}-\frac{H}{\dot\phi}\Bigg(H^2D^{(ab)}_1\cdot Y^aY^b+3HD^{ab}_2\cdot Y^a\dot{Y}^b+3D^{(ab)}_3\cdot \dot{Y}^a\dot{Y}^b+3D^{ab}_4\cdot Y^a\ddot{Y}^b-\frac{3}{2}D^{ab}_5\cdot \dot{Y}^a\ddot{Y}^b\nonumber\\
    &\qquad\qquad+\frac{1}{a^2}D^{ab}_{6}\cdot Y^a\nabla^2 Y^b+\frac{1}{a^2}D^{(ab)}_{7}\cdot \partial_i Y^a\partial^i Y^b+\frac{1}{a^2H}D^{ab}_{8}\cdot \dot{Y}^a\nabla^2 Y^b-\frac{2}{a^2H}D^{ab}_{9}\cdot \partial_i \dot{Y}^a\partial^i Y^b\nonumber\\
    &\qquad\qquad+\frac{1}{a^2H^2}D^{ab}_{10}\cdot \ddot{Y}^a\nabla^2 Y^b+\frac{1}{a^2H^2}D^{(ab)}_{11}\cdot \partial_i \dot{Y}^a\partial^i \dot{Y}^b-\frac{1}{2H^2a^4}D^{(ab)}_{12}\cdot 
    \mathcal{D}_{ij}Y^a\partial^i\partial^jY^b
    \Bigg),\label{eq:S2}
    \end{align}
\end{widetext}
where $A$, $B$, $C$ and $D$ with their respective indices are dimensionless background coefficients whose expressions are given in Appendix~\ref{app:expressions} in terms of $\alpha$ parameters. Note that for each coefficient we have extracted some powers of $H$ and $a$ in order to make them dimensionless, as well as some overall signs and numerical factors in some cases. Importantly, the equations above do not rely on any approximations such as the weak-field limit or quasistatic approximation.
The quadratic coefficients written with brackets in their upper indices, e.g. $A^{(ab)}_1$, are symmetric under an $(a,b)$ swapping. In total, with this notation we have introduced 300 unique coefficients, i.e. 39 linear ($13\times3$), 72 second-order symmetric ($12\times6$) and 189 for the other second-order ones ($21\times9$). However, for the theories considered here, some of these terms are automatically zero. In Tables~\ref{tab:linear_combined} and~\ref{tab:second_order_all} in Appendix~\ref{app:expressions}, we collect all the non-zero linear and second-order coefficients respectively. There, we see that out of the potential 300 terms, only 150 (exactly half) are actually non-zero. Note also that in non-luminal Horndeski theories
this number increases, as more combinations will appear, see e.g. \cite{Kimura:2011dc}.
The expressions above have been computed and organised with the \texttt{Mathematica} package \href{https://github.com/sergisl/xAlpha}{\faGithub \;xAlpha}, which also allows the coefficients to be expressed directly in terms of the $\alpha$ parameters.

\subsection{Standard approximations}
Not all terms in the perturbation equations above are equally relevant in some cosmological settings. In fact, hierarchical structures emerge within some regimes that allow us to safely neglect most contributions. The use of such approximations, along with studies of their regime of validity, is standard practice in large-scale structure studies, see e.g. \cite{Noller:2013wca,Orjuela-Quintana:2023zjm}.

Typically, these simplifications are justified by invoking the \emph{quasi-static approximation} (QSA), often loosely described as neglecting time derivatives in the perturbation equations. Additionally, a weak field limit is typically employed, which exploits the fact that scalar metric perturbations are small on the relevant scales. In addition, one also requires the use of the background equations of motion and, as explained before, the imposition of cosmological mass-scale matchings. Here, we precisely define these assumptions and show how they are used to simplify the perturbation equations.

\subsubsection{Weak field limit}
The weak field approximation dictates the scaling
\begin{equation}
    |\Phi| \sim |\Psi| \sim \mathcal{O}(|u^i|^2) \sim \mathcal{O}(\epsilon) \ll 1
\end{equation}
with $|u^i|$ being the velocity of non-relativistic matter. These velocities are typically around $u/c\sim\mathcal{O}(10^{-3})$ on galactic and cluster scales \cite{Turnbull:2011ty}, meaning that the terms in the perturbation equations which do not contain any spacetime derivatives typically correct other existing terms at an order $\mathcal{O}(10^{-5})$, and can therefore be safely ignored \cite{Winther:2015pta,Fidler:2017pnb}.

This justifies the perturbative treatment of the field amplitudes. However, this hierarchy does not extend to their gradients. The standard Poisson equation relates $\nabla^2\Phi\sim\delta$, where the density contrast $\delta$ can reach $\mathcal{O}(10^2)$ on cluster and $\mathcal{O}(10^{30})$ on Solar System scales. To account for this consistently, we adopt a perturbative scheme for spatial derivatives where, on small scales, $\partial_i\sim\mathcal{O}(\epsilon^{-1/2})$ \cite{Green:2010qy,Green:2011wc,Fidler:2017pnb}. Consequently, double-derivative terms like $\nabla^2\Phi$ act as $\mathcal{O}(1)$ contributions, dominating over single-derivative ($\mathcal{O}(\epsilon^{1/2})$) and non-derivative ($\mathcal{O}(\epsilon)$) terms.
We therefore retain these second-derivative terms for the metric potentials while discarding lower-order spatial gradients.

Importantly, this weak-field limit does not automatically apply to the scalar field $Q$. In many standard linear regimes, the equations of motion do indeed drive $Q$ to the same order of magnitude as the metric potentials. However, naively extending this weak-field limit to $Q$ risks eliminating nonlinear terms responsible for screening. In fact, within screened regimes, the scalar field dynamics can drive the field allowing it to grow. Consequently, we do not enforce $Q\sim \mathcal{O}(\epsilon)$, allowing us to retain nonlinear terms (such as $Q^2$ or $Q\nabla^2 Q$) which might become dynamically important in nonlinear regimes. Finally, we emphasise that the weak-field limit naturally breaks down in strong gravity regimes, such as the vicinity of black holes, where the potentials themselves become non-perturbative. Fully capturing the dynamics of screening in these extreme environments strictly requires full numerical relativity simulations, see e.g. \cite{Cayuso:2026kvg,Cayuso:2024ppe,Lara:2022gof,Bezares:2021dma,Bezares:2021yek,terHaar:2020xxb} in relation to kinetic screening.

\subsubsection{Quasi-static approximation}
The quasi-static approximation (QSA) is built upon the following two assumptions \cite{Boisseau:2000pr,Esposito-Farese:2000pbo,Copeland:2006wr,Tsujikawa:2007gd}:
\begin{itemize}
    \item \textit{Hubble timescale}: The time evolution of perturbation fields is bounded by\footnote{Note that here we use the conformal Hubble parameter, defined as $\mathcal{H} = aH$.}
        \begin{equation}
            |\dot{Y}|\leq \mathcal{H}|Y|.
        \end{equation}
    \item \textit{Sub-horizon scales}: Provided the scalar sound speed squared $c_s^2$ is not too close to zero, perturbations relevant for the evolution of the large-scale structure are characterised by modes deep within the Hubble radius:\footnote{This approximation is often written in Fourier space as $k^2/a^2\gg\mathcal{H}^2$, where $k$ is the 3D wavenumber of the perturbations.}
        \begin{equation}
            |\nabla^2Y|\gg \mathcal{H}^2|Y|.
        \end{equation}
\end{itemize}
Collectively, these conditions define the QSA. In this limit, time derivatives and Hubble-scale terms (of order $\mathcal{H}^2|Y|$, $\mathcal{H}|\dot{Y}|$, and $|\ddot{Y}|$) are neglected in favour of spatial gradients ($a^{-2}|\nabla^2Y|$). Crucially, this hierarchy extends to cross-terms between potentials; for example, we neglect $\mathcal{H}^2\Phi$ even when compared to the spatial Laplacian of the coupled potential, $a^{-2}\nabla^2\Psi$.
Moreover, the QSA is applied identically across all perturbation fields, systematically eliminating their time-derivatives alike. Note, however, that the QSA does not imply that all non-derivative terms (such as $Q^2$) are automatically negligible. While in shift-symmetric theories such terms naturally scale as $\mathcal{H}^2$ (and are therefore discarded), in theories that explicitly break shift-symmetry (such as $f(R)$), the equations introduce effective mass terms dominated by the scalar potential $V_{\phi\phi}$, which are not suppressed by the QSA. See for example how in Eq.~\eqref{eq:D1QQ} $D_1^{QQ}$ is separated into its shift-symmetric and explicit symmetry-breaking components.

\subsection{Effective equations for structure formation}
Having factored out the dimensionful scales to isolate the dimensionless coefficients, we can now easily apply the aforementioned approximations to systematically reduce the perturbation equations.
Following the weak-field and quasi-static counting, all terms in the metric equations are suppressed by $\mathcal{O}(\epsilon)$ relative to the leading-order linear Laplacians, which scale as $\mathcal{O}(1)$.
Interestingly, this means no nonlinear contributions from $\Phi$ and $\Psi$ remain present, a property specific to the luminal Horndeski subclass in contrast to the non-luminal cases \cite{Kimura:2011dc,Crisostomi:2017lbg}.
In particular, the metric $(0,0)$ equation evaluated on the background becomes
\begin{align}
    2\nabla^2\Psi+\alpha_B\nabla^2 Q=a^2\tilde{\rho}_m\delta,
    \label{eq:poisson}
\end{align}
while the traceless part of the metric $(i,j)$ equation becomes\footnote{\red{We do not display the trace part of the metric equations since we ultimately model matter as a pressureless perfect fluid ($p_m=0$). It can be found, nonetheless, in the companion notebook in \href{https://github.com/sergisl/xAlpha}{\faGithub \;xAlpha}.}}
\begin{align}
    \nabla^2(\Psi-\Phi-\alpha_MQ)=0.
    \label{eq:lensing}
\end{align}
Note that in the limit where $\alpha_M=0$, as in e.g. the cubic Galileon, the potentials satisfy $\nabla^2\Psi = \nabla^2\Phi$. This leads to $\Psi = \Phi$, which is the defining property of `no-slip' gravity theories \cite{Linder:2018jil}.
The $(0,i)$ metric equation is heavily suppressed under the QSA.
Finally, the scalar equation becomes\footnote{Note that \cite{Kimura:2011dc} uses the definition $\mathcal{Q}^{(2)}\equiv\mathcal{D}_{ij}Q\partial^i\partial^jQ$.}
\begin{align}
    &\left[\gamma_B-\gamma_E-2(\alpha_B-\alpha_M)\right]\nabla^2Q+\alpha_B\nabla^2\Phi+2\alpha_M\nabla^2\Psi\nonumber\\
    &\qquad -a^2M^2 Q-a^2M_{\rm nl}^{2} Q^2+\kappa_{-}Q\nabla^2Q+\kappa_{+}(\partial_iQ)^2\nonumber\\
    &\qquad-\mathcal{H}^{-2}(\alpha_B+\alpha_M)\mathcal{D}_{ij}Q\partial^i\partial^jQ=0,
    \label{eq:scalar-pert-eom1}
\end{align}
where we have defined the following parameters:
\begin{align}
    M^2&\equiv -H^2D^Q_1, &
    M_{\rm nl}^{2}&\equiv -H^2D^{QQ}_1,\nonumber\\
    \kappa_{-}&\equiv D^{QQ}_6, &
    \kappa_{+}&\equiv D^{QQ}_7,
    \label{eq:relabel-coeffs}
\end{align}
and recall that $D_1^Q$ \eqref{eq:D1Q}, $D_1^{QQ}$ \eqref{eq:D1QQ}, $D_6^{QQ}$ \eqref{eq:D6QQ} and $D_7^{QQ}$ \eqref{eq:D7QQ} are defined in the referred equations in Appendix \ref{app:expressions}. Note that we have absorbed $H^2$ into the definitions of mass terms, thus making them dimensionful quantities. The choice of $\{-,+\}$ subindices for the $\kappa$ parameters will become evident in Section~\ref{sec:phaedrus}.
We observe that nonlinearities manifest solely in the scalar equation and involve purely self-interactions of $Q$. Under standard perturbative counting, one might expect only the four-derivative term $\mathcal{D}_{ij}Q\partial^i\partial^jQ \sim \mathcal{O}(1)$ to survive, as it naturally competes with the linear Laplacian. However, to maintain a fully generic framework capable of describing the primary screening phenomenologies, we must also retain specific $\mathcal{O}(\epsilon)$ scalar nonlinearities. In theories exhibiting Chameleon screening, the effective mass terms are dynamically enhanced in high-density environments. Similarly, in models with non-canonical kinetic terms, the $Q\nabla^2Q$ and $(\partial_iQ)^2$ terms might also be promoted.\footnote{As mentioned in the introduction, we leave the inclusion of the symmetron and $k$-mouflage mechanisms for future work, as they require third-order perturbations.}
Recall that different choices for the scalar variable (i.e. $Q$, $v_X$ or $\delta\phi$) yield different expressions for the coefficients. As an example, we detail the translation of the linear mass term $M^2$ across these conventions in Appendix~\ref{app:conversions}.

The effective equations derived above serve as the starting point to compute theoretical predictions for key cosmological probes, including e.g. the matter power spectrum \cite{Winther:2015wla, Wright:2022krq}, weak lensing statistics \cite{Barreira:2015fpa,Barreira:2015vra}, the integrated Sachs-Wolfe effect \cite{Seraille:2024beb}, and the growth of structure probed via redshift-space distortions \cite{Bose:2016qun}.
Here, we are specifically interested in how different screening mechanisms are encoded within the nonlinear terms in the scalar equation. In order to do so, we i) use the modified Poisson equation \eqref{eq:poisson} to solve for $\Psi$, ii) substitute that into the modified gravitational slip equation \eqref{eq:lensing}, iii) solve the latter for $\Phi$, and iv) substitute both solutions into the scalar equation \eqref{eq:scalar-pert-eom1}, obtaining
\begin{widetext}
\begin{gather}
\Gamma\nabla^2Q-a^2Q\big[M^2+\underbrace{M_{\rm nl}^2Q}_\text{Chameleon}\big]+\underbrace{\kappa_{-}Q\nabla^2Q+\kappa_{+}(\partial_iQ)^2}_\text{Phaedrus}-\underbrace{\mathcal{H}^{-2}(\alpha_B+\alpha_M)\mathcal{D}_{ij}Q\partial^i\partial^jQ}_\text{Vainshtein}
=-\frac{1}{2}(\alpha_B+2\alpha_M)a^2\tilde{\rho}_m\delta,
\label{eq:scalar-pert-eom}
\end{gather}
\end{widetext}
where $\Gamma \equiv D c_s^2$ represents the effective linear spatial kinetic term, with $c_s^2$ being the standard scalar sound speed and $D$ the no-ghost parameter (see Appendix~\ref{app:cov-eoms} for their exact expressions).
Eq.~\eqref{eq:scalar-pert-eom} is a quadratic nonlinear differential equation, whose general analytical solution is not known. Hence, in order to solve this unified screening equation, we will resort to several simplifying assumptions and numerical methods. As highlighted by the under-brackets in Eq.~\eqref{eq:scalar-pert-eom}, each of the retained nonlinear terms dictates a distinct screening phenomenology. The term $\mathcal{D}_{ij}Q\partial^i\partial^jQ$ drives the Vainshtein mechanism by suppressing the scalar field through second-order spatial derivatives. The $M_{\rm nl}^2$ Chameleon term provides screening by dynamically increasing the effective mass of the field in dense regions.
Finally, the $\kappa_{-}$ and $\kappa_{+}$ terms source \red{what we identify here as the candidate Phaedrus screening regime, in which field-dependent non-canonical kinetic interactions dominate the field's evolution and suppress the fifth force.\footnote{\red{Because matter is minimally coupled to the Jordan-frame metric, test masses follow exact geodesics of the physical spacetime. Therefore, no literal non-geodesic `fifth force' acts on matter. Throughout this work, we use the term `fifth force' to denote the scalar-induced modification to the metric potentials relative to the standard GR prediction.}} As we discuss in Sec.~\ref{sec:phaedrus-viability}, whether this regime constitutes a genuinely independent physical mechanism requires further validation.}

\red{We emphasise that Eq.~\eqref{eq:scalar-pert-eom} captures the nonlinear operators visible through second-order perturbations. It therefore does not capture mechanisms controlled by higher powers of the scalar or its kinetic term (such as $K$-mouflage or the Symmetron) but rather a systematic classification of the second-order operators and the screening behaviour they generate. Nonetheless, by extending this formulation to higher perturbative orders, such remaining mechanisms could also be incorporated.}

Before examining these mechanisms individually, it is worth highlighting the universal nature of the source term in Eq.~\eqref{eq:scalar-pert-eom}. The coupling to the local matter overdensity $\delta$ on the right-hand side is strictly governed by the combination $(\alpha_B + 2\alpha_M)$. The condition $\alpha_B + 2\alpha_M = 0$ therefore plays a fundamental role across this entire framework: it entirely decouples the scalar field from local matter perturbations. In this limit, the source term identically vanishes, meaning no local scalar profile is generated. Consequently, the fifth force is never sourced, none of the aforementioned screening mechanisms are required to operate, and local deviations from GR strictly vanish.\red{ Equivalently, in this limit the effective gravitational coupling reduces to its GR value, $G_{\rm eff}=G_N$, as follows directly from Eq.~\eqref{eq:Geff}.}

\section{Screening mechanisms}\label{sec:screening}
Having derived the generic equation governing the scalar nonlinear dynamics, let us now see how one can extract screening phenomenology from it. We do so by considering a spherically symmetric overdensity embedded in a cosmological background,\footnote{\red{Throughout this section we work at a fixed cosmological background, treating the analysis as a snapshot in cosmic time, where the background couplings ($\alpha_B$, $\alpha_M$, $\Gamma$, $H$, etc.) and the mean density are held constant, and only the radial dependence of the profiles is solved for.}} with radial coordinate \cite{Kimura:2011dc}
\begin{equation}
r = a(t)\sqrt{\delta_{ij}x^i x^j}.
\end{equation}
Our focus is on scales well within the Hubble radius, $rH \ll 1$, where the background metric can be approximated as
\begin{equation}
ds^2 \simeq -dt^2 + dr^2 + r^2 d\Omega^2,
\end{equation}
with $d\Omega^2$ denoting the line element of the unit two-sphere. Spherical symmetry allows the Laplacian and related derivatives to be written as
\begin{align}
a^{-2}\nabla^2 Q &= r^{-2} (r^2 Q')', \\
a^{-4} \mathcal{D}_{ij}Q\partial^i\partial^jQ &= 2 r^{-2} [r (Q')^2]',\\
a^{-2}(\partial_iQ)^2&=(Q')^2
\end{align}
where a prime denotes differentiation with respect to $r$. This results in the nonlinear scalar equation
\begin{align}
    &\Gamma\frac{(r^2Q')'}{r^2} + \mathcal{C} + \mathcal{P} + \mathcal{V} = -\frac{1}{2}(\alpha_B+2\alpha_M)\tilde{\rho}_m\delta,
    \label{eq:scalar-pert-eom2}
\end{align}
where we have defined the functions
\begin{align}
    \mathcal{C}(Q) &= -Q(M^2+M_{\rm nl}^2Q), \nonumber\\
    \mathcal{P}(Q) &= \kappa_{-}\frac{1}{r^2}Q(r^2Q')' + \kappa_{+}(Q')^2, \nonumber\\
    \mathcal{V}(Q) &= -\frac{2}{r^2 H^2}(\alpha_B+\alpha_M)[r(Q')^2]',
\end{align}
each corresponding to different mechanisms suppressing fifth forces.

In the following subsections, we shall define $Q_\mathcal{C}$, $Q_\mathcal{P}$ and $Q_\mathcal{V}$ as the corresponding solutions to Eq.~\eqref{eq:scalar-pert-eom2} where we solely include either $\mathcal{C}$, $\mathcal{P}$ or $\mathcal{V}$ and ignore the rest.

\begin{figure}
    \centering
    \includegraphics[width=\linewidth]{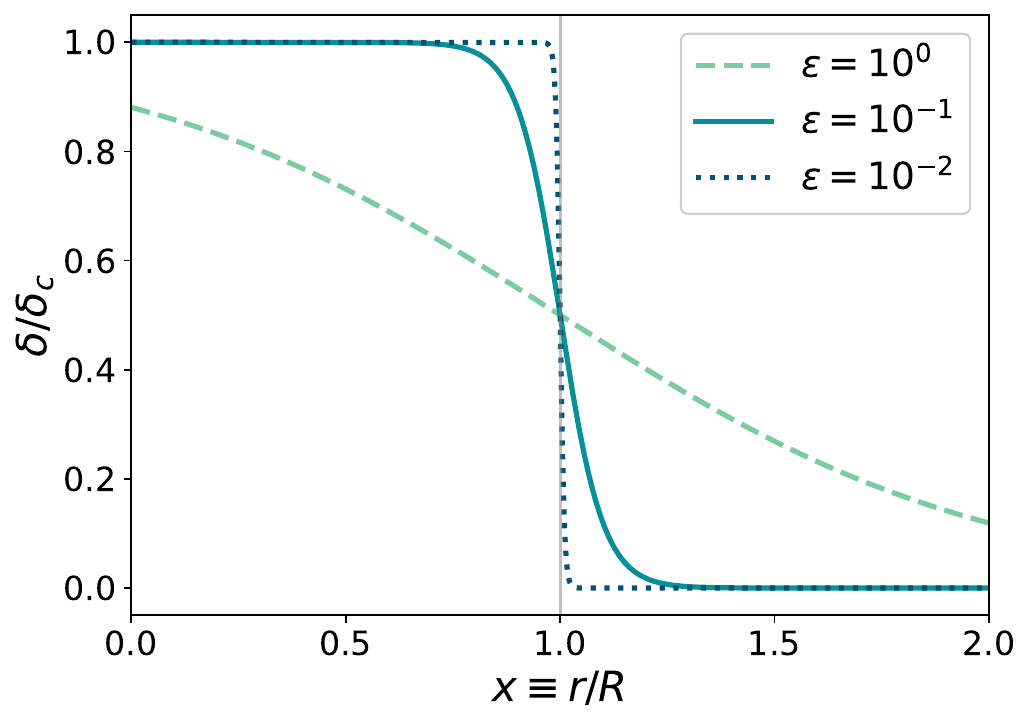}
    \caption{Smoothed normalised top-hat density profile~\eqref{eq:eps-def} as a function of the normalised radial coordinate $x\equiv r/R$ for different edge parameters $\epsilon$.}
    \label{fig:top-hat-density}
\end{figure}

We model the local source density with a smoothed top-hat profile described by
\begin{align}
\delta(r)
&= s(r)\,\delta_c + \bigl[1 - s(r)\bigr]\delta_\infty,\nonumber
\\
s(r) &= \frac12\!\left[1 - \tanh\!\left(\frac{\frac{r}{R}-1}{\varepsilon}\right)\right],\label{eq:eps-def}
\end{align}
where $\delta_c$ and $\delta_\infty$ are the (constant) densities inside and outside the source of radius $R$, respectively. The dimensionless parameter $\varepsilon>0$ controls the width of the transition region between the interior and exterior values, as shown in Fig.~\ref{fig:top-hat-density}. For $\varepsilon\ll 1$ the profile approaches a sharp top-hat with a very narrow boundary, while for larger $\varepsilon$ the transition is smoother and extends over a wider radial interval. We also define the integrated total mass as
\begin{align}
\mu &\equiv \frac{\tilde{\rho}_{\rm m}}{2} \int^r \delta(r')\, r'^2\, dr'.
\label{eq:mu}
\end{align}

Finally, note that, while here we assume spherically symmetric sources, cylindrical or planar solutions, such as those studied in \cite{Bloomfield:2014zfa}, are also of cosmological interest.

\subsection{The Linear Regime}
Before investigating the specific nonlinear screening mechanisms, we first establish the universal behaviour of the scalar field at large distances. Far from the source, the field amplitude $Q$ and its gradients are small, allowing us to linearise the full master equation~\eqref{eq:scalar-pert-eom2}
\begin{equation}
\frac{(r^2 Q')'}{r^2} - m_{\rm eff}^2 Q \approx -\frac{1}{2\Gamma}(\alpha_B+2\alpha_M)\tilde{\rho}_m\delta,
\label{eq:linear-Q}
\end{equation}
where we have redefined the linear effective mass of the scalar perturbation as $m_{\rm eff}^2 \equiv M^2 / \Gamma$. The solution outside a source of radius $R$ and mass $\mu$ is given by the Yukawa profile
\begin{equation}
Q_{\rm Yuk}(r) = \frac{(\alpha_B+2\alpha_M)\mu}{\Gamma r} e^{-m_{\rm eff} r}.
\label{eq:Q_Yukawa}
\end{equation}
This exponential suppression is the hallmark of massive scalar theories. In the context of screening, this linear suppression is the starting point for the Chameleon mechanism, where the nonlinearities act to increase the effective mass $m_{\rm eff}^2$ in dense regions, further shrinking the force range. Alternatively, for theories relying on kinetic screening (such as Vainshtein or $K$-mouflage), the field is typically assumed to be light or massless on astrophysical scales ($M^2 \ll \Gamma / r^2$). Taking the limit $m_{\rm eff} \to 0$ in Eq.~\eqref{eq:Q_Yukawa}, we recover the standard Newtonian $1/r$ decay:
\begin{equation}
Q_{\rm lin}(r) = \frac{(\alpha_B+2\alpha_M)\mu}{\Gamma r}.
\label{eq:Q_Newtonian}
\end{equation}
This solution serves as the universal asymptote for all kinetic screening mechanisms.

The \textit{screening radius} $r_\star$ for any such mechanism is formally defined as the characteristic distance at which the magnitude of the leading-order nonlinear term equals that of the linear term, marking the transition into the screened regime~\cite{Brax:2013ida,Babichev:2013usa,Joyce:2014kja,Brax:2014wla}.
To quantify these deviations systematically for kinetic screening types, we introduce the \textit{screening efficiency slope} parameter, $n$, defined as the logarithmic derivative of the scalar flux
\begin{equation}
n(r) \equiv \frac{d \ln |r^2 Q'(r)|}{d \ln r}.
\label{eq:n_definition}
\end{equation}
Physically, $n$ measures how effectively the fifth force is suppressed relative to gravity as we approach the source. Since the fifth force scales as $F_5 \propto Q' \propto r^{n-2}$, and Newtonian gravity scales as $F_N \propto r^{-2}$, their ratio evolves as:
\begin{equation}
\frac{F_5}{F_N} \propto r^n.
\end{equation}
This allows us to classify mechanisms by their `screening efficiency' parameter $n$ in the nonlinear regime ($r < r_\star$). If $n=0$, the fifth force remains unscreened, following the standard Newtonian inverse-square law ($F_5 \propto r^{-2}$) such that the ratio $F_5/F_N$ is constant across all scales. Conversely, for screened models ($n > 0$), the nonlinear terms suppress the growth of the fifth force at small radii, driving $F_5/F_N \to 0$ deep inside the screening radius. 
This parameter provides a unified metric to compare the kinetic screening mechanisms derived in the following sections, as summarised in Table~\ref{tab:screening_slopes}.

\begin{table}[h]
\centering
\renewcommand{\arraystretch}{1.2} 
\begin{tabular}{l c c c}
\toprule
Mechanism & Force Profile ($Q'$) & Efficiency ($n$) & \red{Radius ($r_\star$)} \\
\midrule
Vainshtein & $\propto r^{-1/2}$ & $1.5$ & \red{$\propto \mu^{1/3}$} \\
$K$-mouflage & $\propto r^{-2/3}$ & $4/3 \approx 1.33$ & \red{$\propto \mu^{1/2}$} \\
Phaedrus & $\propto r^{-1} \to r^{-2}$ & $0 \to 1$ & \red{$\propto \mu$} \\
Unscreened & $\propto r^{-2}$ & $0$ & \red{--} \\
\bottomrule
\end{tabular}
\caption{
Classification of kinetic screening mechanisms by their screening efficiency parameter $n$, with higher $n$ resulting in more efficient suppression of the fifth force.
The Vainshtein entry, producing the most efficient screening, corresponds to the cubic Galileon, the unique luminal operator in this class; quartic and quintic (non-luminal) interactions would yield stronger screening with $n=2$ and $n=2.25$, respectively. For $K$-mouflage, we assume a canonical scaling of $K \propto X^m$ with $m=2$; for a general power law $m$, the efficiency scales as $n=4(m-1)/(2m-1)$. For the Phaedrus effect identified in this paper, the efficiency ranges from $0$ (unscreened) to $1$ depending on the dominant nonlinear terms. \red{The final column gives the scaling of the screening radius with source mass $\mu$.} Finally, also note that different force profiles are obtained for non-spherically symmetric sources \cite{Bloomfield:2014zfa}.
}
\label{tab:screening_slopes}
\end{table}

\red{The force ratio $F_5/F_N$ introduced above is, moreover, directly tied to a physical observable. As has been mentioned, the scalar $Q$ is itself convention-dependent: the equivalent choices $Q$, $v_X$ and $\delta\phi$ (see Appendix~\ref{app:conversions}) rescale $Q'$ by time-dependent background factors. To show that the suppression of $Q'$ carries a physical meaning, we substitute the scalar field back into the metric equations and derive the dynamical potential $\Phi$ that governs the geodesics of non-relativistic matter. Eliminating $\Psi$ between Eqs.~\eqref{eq:poisson} and~\eqref{eq:lensing} gives
\begin{align}
    \nabla^2\Phi = \frac12 a^2\tilde{\rho}_m\delta
    -\Big(\frac{\alpha_B}{2}+\alpha_M\Big)\nabla^2 Q ,
    \label{eq:Phi-poisson}
\end{align}
and, integrating once over the spherical source, the total radial acceleration splits into the Newtonian term plus a scalar fifth force contribution,
\begin{align}
    \Phi'(r) = \Phi'_{\rm GR}
    -\Big(\frac{\alpha_B}{2}+\alpha_M\Big)Q'(r),
    \label{eq:fifth-force}
\end{align}
where $\Phi'_{\rm GR}\equiv \mu(r)/r^2$. 
The effective gravitational coupling felt by matter, $G_{\rm eff}/G_N\equiv\Phi'/\Phi'_{\rm GR}=1+F_5/F_N$, then follows on inserting the linear flux $Q'_{\rm lin}$ from Eq.~\eqref{eq:Q_Newtonian},
\begin{align}
    \frac{G_{\rm eff}}{G_N}
    = 1 + \frac{(\alpha_B+2\alpha_M)^2}{2\,\Gamma}\,
    \frac{Q'(r)}{Q'_{\rm lin}(r)} .
    \label{eq:Geff}
\end{align}
Far from the source $Q'\to Q'_{\rm lin}$ and one recovers the standard linear enhancement. Alternatively, wherever the nonlinear terms screen the flux, $Q'/Q'_{\rm lin}\to0$ and $G_{\rm eff}/G_N\to1$, so the dynamical potential becomes indistinguishable from its GR counterpart. Since $G_{\rm eff}/G_N$ is defined as the ratio of the physical metric acceleration $\Phi'$ to its GR value, it is independent of the convention used for the scalar perturbation. In other words, although Eq.~\eqref{eq:Geff} is written through $Q'$, the ratio $Q'/Q'_{\rm lin}$ is insensitive to that choice. We evaluate this ratio explicitly for the screened numerical solutions in Sec.~\ref{sec:phaedrus} (bottom panel of Fig.~\ref{fig:phaedrus_diagnostics}).}

\subsection{Vainshtein}
\begin{figure}
    \centering
    \includegraphics[width=\linewidth]{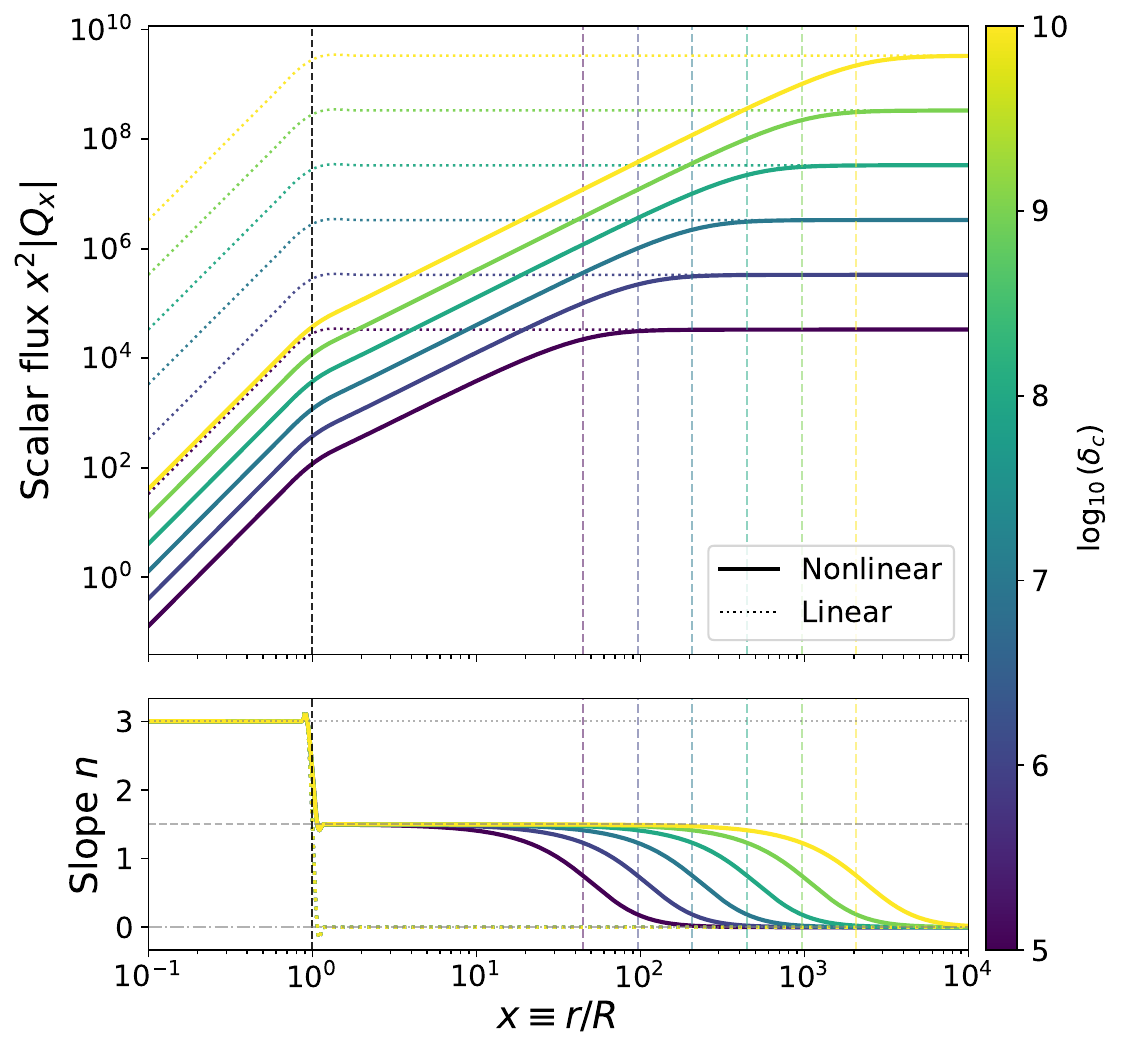}
    \caption{
    Numerical solutions of the Vainshtein mechanism as a function of the normalised radial coordinate for a spherical top-hat density (surface at $x=1$).
    \textbf{Upper panel:} The scalar flux $x^{2}\lvert Q_x\rvert$, where $Q_x \equiv \mathrm{d}Q/\mathrm{d}x$. Solid curves represent the full nonlinear solutions, while dotted curves indicate the corresponding unscreened linear solutions, highlighting the strong suppression of the fifth force deep inside the screened region. 
    \textbf{Lower panel:} The screening efficiency slope $n$~\eqref{eq:n_definition}, transitioning from the source interior ($n=3$ for $x < 1$) to the analytical Vainshtein plateau ($n=1.5$), before eventually decaying to the unscreened limit ($n=0$). 
    Different colours correspond to varying density amplitudes, demonstrating that denser sources possess a larger Vainshtein radius $r_\mathcal{V}$ (marked for each solution by the corresponding vertical dashed lines).
    }
    \label{fig:vainshtein_diagnostics}
\end{figure}
The Vainshtein mechanism, originally proposed in~\cite{Vainshtein:1972sx}, relies on nonlinear derivative interactions to dynamically suppress scalar fifth forces in high-density environments, effectively restoring GR on local scales.\footnote{For a comprehensive pedagogical review see~\cite{Babichev:2013usa}, and for generalising the Vainshtein mechanism into a broader class of scalar field theories see~\cite{Nicolis:2008in}; for more recent investigations detailing its cosmological applications, N-body implementations, and impact on structure formation, see e.g.~\cite{Barreira:2013eea,Li:2013tda,Barreira:2013xea,Winther:2015wla, Dima:2017pwp,Taniguchi:2022djn,Wright:2022krq}.}
Isolating the Vainshtein operator $\mathcal{V}$ in Eq.~\eqref{eq:scalar-pert-eom2} and neglecting for now the other terms, the first integral of the equation of motion yields a quadratic algebraic equation for the field gradient quantity $Q'/r$:
\begin{equation}
\frac{2(\alpha_B+\alpha_M)}{H^2\Gamma} \left( \frac{Q_\mathcal{V}'}{r} \right)^2 - \frac{Q_\mathcal{V}'}{r} = \frac{(\alpha_B+2\alpha_M)}{\Gamma} \frac{\mu}{r^3}.
\label{eq:QVainsh}
\end{equation}
Hence, we see that the nonlinear Vainshtein interaction is sourced entirely by the cubic Galileon coefficient $(\alpha_B+\alpha_M)\propto G_{3X}$.
Eq.~\eqref{eq:QVainsh} admits the exact solution
\begin{equation}
\frac{Q_\mathcal{V}'}{r} = -\frac{H^2\Gamma}{4(\alpha_B+\alpha_M)}
\left[
\sqrt{1 + \frac{2r_\mathcal{V}^3}{r^3}} - 1
\right],
\label{eq:Q'}
\end{equation}
where the negative branch of the square root is chosen to ensure the gradient vanishes at infinity. Here, we have defined the Vainshtein radius $r_\mathcal{V}$, characterising the scale of the screened region, as
\begin{equation}
r_\mathcal{V}^3 \equiv \frac{4(\alpha_B+\alpha_M)(\alpha_B+2\alpha_M)}{\Gamma^2} \frac{\mu}{H^2}.
\label{eq:rV}
\end{equation}
Outside this radius ($r \gg r_\mathcal{V}$), the term under the square root is negligible, and expanding Eq.~\eqref{eq:Q'} recovers the linear Newtonian solution in Eq.~\eqref{eq:Q_Newtonian}. However, inside the Vainshtein radius ($r < r_\mathcal{V}$), the nonlinear derivative interactions become dominant and the gradient scales as $Q_\mathcal{V}' \propto r^{-1/2}$, hence with a screening efficiency of $n=1.5$. Integrating Eq.~\eqref{eq:Q'} in this limit yields the screened field potential
\begin{equation}
 Q_{\mathcal{V}}(r) \simeq Q_0 - \frac{\Gamma H^2}{2(\alpha_B+\alpha_M)}\sqrt{2r_\mathcal{V}^3r},
\end{equation}
where $Q_0$ is a positive integration constant determined by matching this interior profile to the exterior linear solution at $r \approx r_\mathcal{V}$.

The numerically obtained radial dependence of the field from the interior of the source to the asymptotic linear regime is visualised in Fig.~\ref{fig:vainshtein_diagnostics}. The upper panel shows how the nonlinear suppression strongly restricts the growth of the scalar flux relative to the unscreened Newtonian expectation within the Vainshtein regime. Here, the numerical solution perfectly captures the analytical plateau at $n=1.5$. In fact, we have confirmed that the numerical solutions agree with the analytical expressions to great accuracy. Beyond this, the nonlinearities decay, and the field smoothly transitions to the unscreened linear regime ($n=0$). As expected from Eq.~\eqref{eq:rV}, increasing the source density $\delta_c$ extends the Vainshtein radius further outward, increasing the spatial reach of the screening mechanism.

Note that for this analytical screened solution to be physically well-defined, we require the Vainshtein radius to be a real, positive quantity ($r_\mathcal{V}^3 > 0$). Assuming a standard overdensity ($\mu > 0$), this imposes the condition:
\begin{equation}
(\alpha_B+\alpha_M)(\alpha_B+2\alpha_M) > 0.
\end{equation}
Under this condition, substituting the screened solution back into the metric field equations confirms that in the short-distance limit we recover the General Relativity prediction $\Phi \approx \Psi$, ensuring the theory remains compatible with Solar System tests.
However, the behaviour of this solution changes in underdense regions ($\delta_c < 0 \implies \mu < 0$), where the quantity $r_\mathcal{V}^3$ becomes negative. In mild underdensities, the term inside the square root of Eq.~\eqref{eq:Q'} can remain positive, but the nonlinear interactions are too weak to dominate the linear terms. Consequently, the Vainshtein mechanism operates highly inefficiently, leaving the scalar fifth force largely unscreened and rendering these regions ideal for observing deviations from GR~\cite{Falck:2014jwa, Barreira:2015vra, Baker:2018mnu}.
Conversely, in deep cosmic voids where the density drops below a critical threshold (such that $2|r_\mathcal{V}^3|/r^3 > 1$), the term inside the square root becomes strictly negative, yielding an imaginary field gradient.
Numerical investigations relaxing the QSA encounter this identical pathology, suggesting that it represents a genuine theoretical instability inherent to certain models~\cite{Winther:2015pta}. Note that in those cases simulations break down and the standard practice is to set the scalar to a constant value within deep voids, see e.g.~\cite{Winther:2015wla,Wright:2022krq}.
The requirement that a theory remains stable and admits real solutions inside realistic cosmic voids therefore places strict constraints on the viable parameter space of these models~\cite{Takadera:2025ehm}.

\subsection{Chameleon}
\begin{figure}
    \centering
    \includegraphics[width=\linewidth]{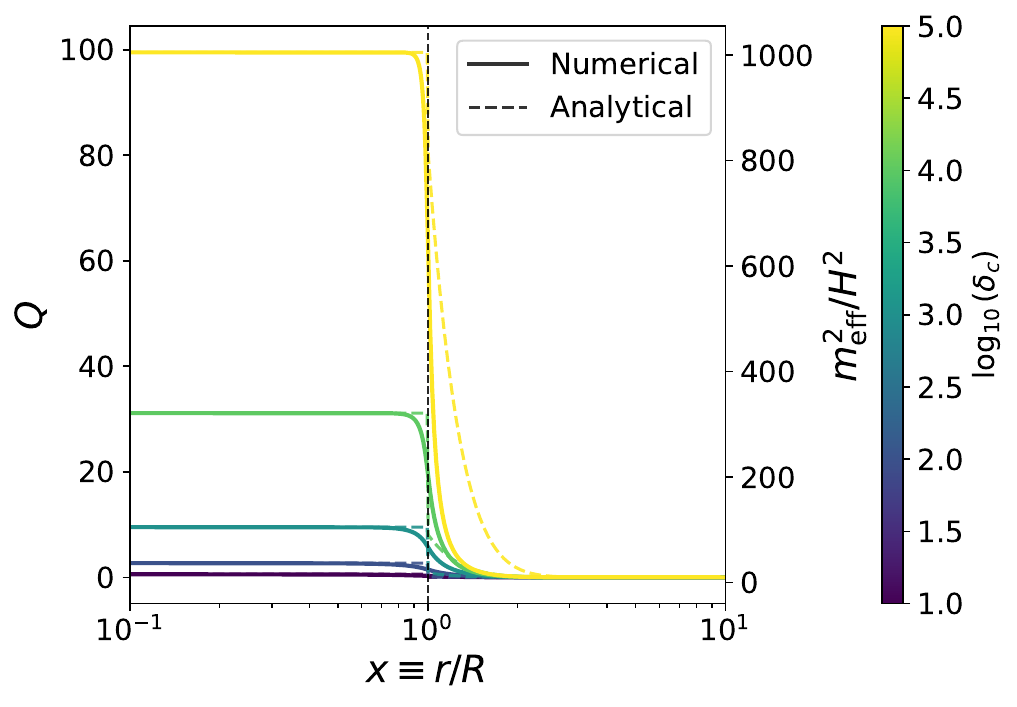}
    \caption{Scalar field $Q$ (left axis) and effective Chameleon mass $m_{\rm eff}^2/H^2$ (right axis) as a function of the normalised radial coordinate. Both quantities share identical profile shapes due to the relation in Eq.~\eqref{eq:meff}. Solid curves represent the full nonlinear numerical solutions for $Q$, while dashed curves indicate the corresponding analytical approximations in Eq.~\eqref{eq:thin-shell-sol}. Inside the source ($x < 1$), the effective mass is driven to a large constant value, dynamically locking the scalar field to the minimum of its effective potential. At the surface ($x=1$), the mass drops sharply, freeing the field to roll in the exterior. Different colours correspond to varying density amplitudes, demonstrating that denser sources generate a larger interior effective mass.}
    \label{fig:chameleon_Q_meff}
\end{figure}
\begin{figure}[ht]
    \centering
    \includegraphics[width=\linewidth]{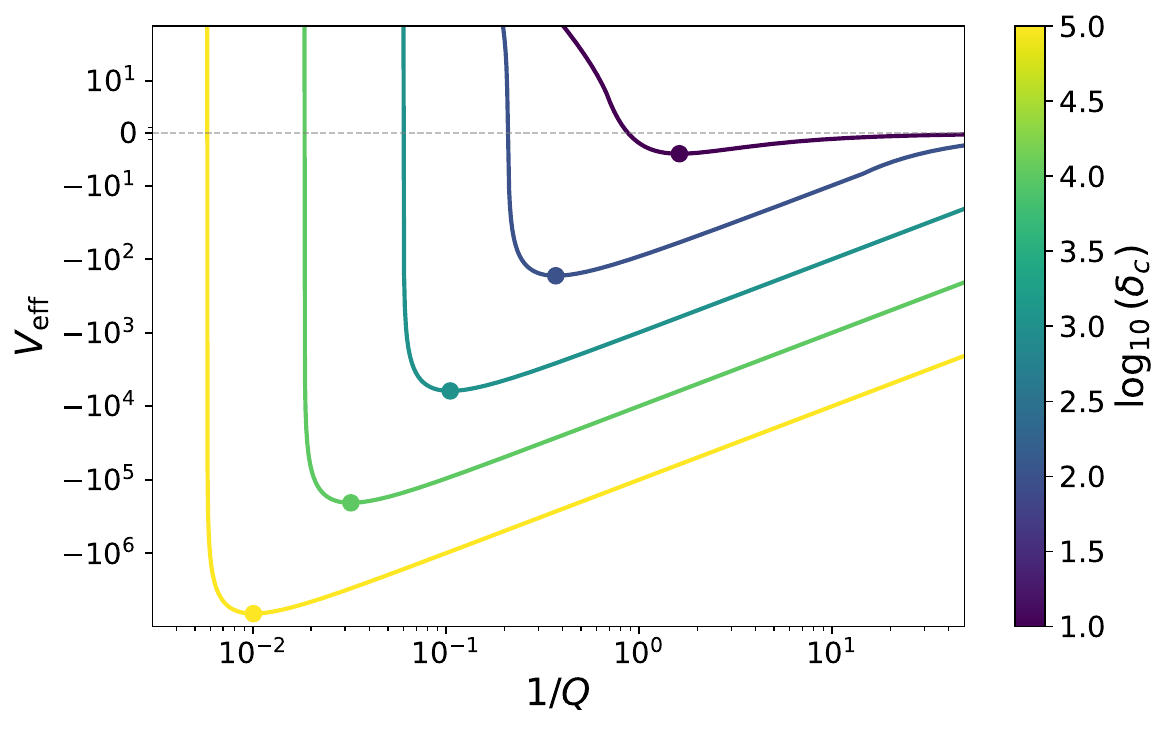}
    \caption{Perturbative Chameleon effective potential $V_{\rm eff}$ as a function of $1/Q$. Solid markers denote the potential minimum, which is observed to shift for varying density amplitudes (in different colours).
    As density increases, the minimum is driven deep into the steep potential wall, dynamically generating the large effective mass required for screening.
    }
    \label{fig:chameleon_potential_inverse}
\end{figure}
\begin{figure}
    \centering
    \includegraphics[width=\linewidth]{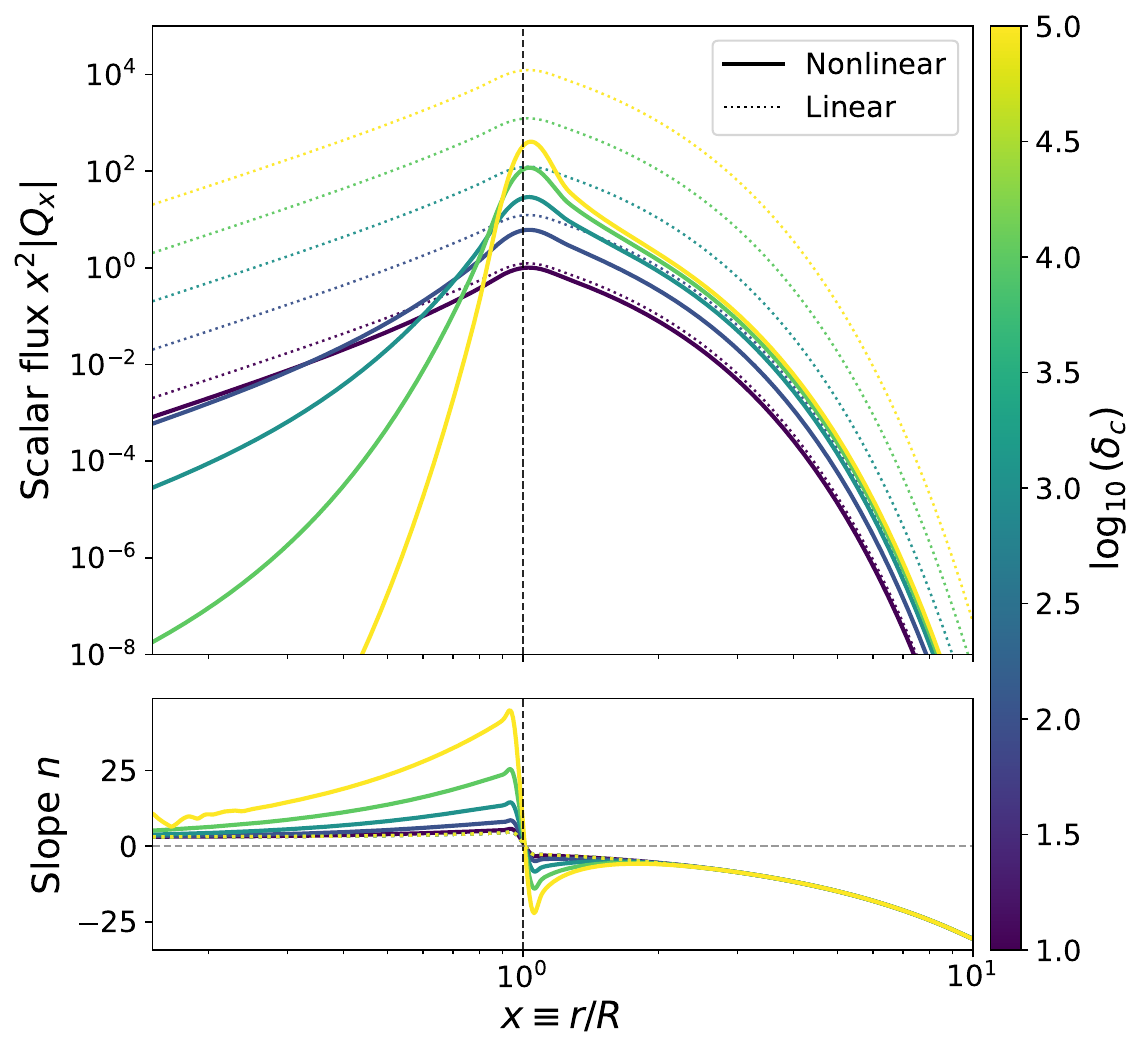}
    \caption{
    Numerical solutions of the Chameleon mechanism as a function of the normalised radial coordinate for a spherical top-hat density (surface at $x=1$). 
    \textbf{Upper panel:} The scalar flux $x^{2}\lvert Q_x\rvert$. Solid curves represent the full nonlinear solutions, while dotted curves indicate the corresponding unscreened linear solutions, highlighting the strong suppression of the fifth force deep inside the source ($x < 1$) due to the thin-shell effect. 
    \textbf{Lower panel:} The screening efficiency slope $n$~\eqref{eq:n_definition}, illustrating the rapid variation of the field across the active thin shell near the surface, before eventually decaying to the exponential Yukawa profile. 
    Different colours correspond to varying density amplitudes, demonstrating that denser sources exhibit a more heavily suppressed interior and a thinner active shell.
    }
    \label{fig:chameleon}
\end{figure}
We now turn to the regime where the Chameleon mechanism dominates over other nonlinear effects. Originally proposed in~\cite{Khoury:2003aq,Khoury:2003rn} and extensively developed for cosmological settings in~\cite{Brax:2004qh}, this mechanism dynamically screens scalar fifth forces by coupling the field to local matter, rendering its effective mass fundamentally density-dependent. In high-density environments, such as the Solar System, the scalar field acquires a large mass, exponentially suppressing its interaction range. Conversely, in low-density cosmological voids, the field remains light, allowing the fifth force to propagate over macroscopic scales and modify structure formation.\footnote{For comprehensive reviews on Chameleon screening and cosmological tests, see e.g.,~\cite{Lombriser:2014dua,Burrage:2017qrf}.}

While the Chameleon literature is frequently formulated in the Einstein frame, related to the Jordan frame via the conformal transformation $g^J_{\mu\nu}=A^2(\phi)\,g^E_{\mu\nu}$, we retain the Jordan frame formulation throughout this work. This choice is physically motivated by the fact that the Jordan frame is the natural observational frame, as common matter fields follow the geodesics of $g^J_{\mu\nu}$. Note that it has been shown that physical predictions in the weak-field, non-relativistic limit remain identical in both frames~\cite{Hui:2009kc,Burrage:2017qrf,Copeland:2021qby}.\footnote{For a conformal factor $A^2 \approx 1 - 2\alpha\phi$, the metric potentials transform as
\begin{align}
    \tilde{\Phi}_E &= \Phi_J - \alpha\phi, & \tilde{\Psi}_E &= \Psi_J + \alpha\phi,
\end{align}
which leaves the lensing potential invariant: $\tilde{\Phi}_E+\tilde{\Psi}_E = \Phi_J+\Psi_J$.}
Our Jordan-frame perturbative Chameleon equation takes the form
\begin{equation}
\frac{1}{r^2} (r^2 Q_\mathcal{C}')' - m_{\rm eff}^2 Q_\mathcal{C} = -\frac{1}{2\Gamma}(\alpha_B+2\alpha_M)\tilde{\rho}_m\delta,
\label{eq:QCeq}
\end{equation}
where now the effective mass term becomes a nonlinear quantity given by
\begin{align}
    m_{\rm eff}^2&=\frac{1}{\Gamma}(M^2+M^2_{\rm nl}Q_\mathcal{C}).
    \label{eq:meff}
\end{align}
Exact closed-form solutions to Eq.~\eqref{eq:QCeq} are not known, but analytic control is possible in the limits $r\ll R$ and $r\gg R$, where $R$ is the source surface.
The local dynamics of the field are governed by its effective potential, defined such that the equation of motion takes the standard form $\nabla^2 Q_\mathcal{C} = V_{\rm eff}'(Q_\mathcal{C})$, hence giving
\begin{equation}
    V_{\rm eff}(Q_\mathcal{C}) = \frac{Q_\mathcal{C}}{2\Gamma} \left[ M^2 Q_\mathcal{C} + \frac{2}{3}M_{\rm nl}^2 Q_\mathcal{C}^2 - (\alpha_B+2\alpha_M)\tilde{\rho}_m\delta \right],
    \label{eq:Veff}
\end{equation}
which is plotted in Fig.~\ref{fig:chameleon_potential_inverse}.
Deep inside the object, the scalar quickly relaxes to the minimum of this local effective potential. In this high-density regime, spatial gradients are subdominant ($\nabla^2 Q_\mathcal{C} \approx 0$), and the minimisation condition $V_{\rm eff}'(Q_\mathcal{C}) = 0$ reduces the dynamics to an algebraic balance between the linear and nonlinear mass terms,
\begin{align}
    M^2 Q_\mathcal{C} + M_{\rm nl}^2 Q_\mathcal{C}^2 \simeq \frac{1}{2}(\alpha_B+2\alpha_M)\tilde{\rho}_m\delta.
\end{align}
Solving this quadratic equation and picking the branch continuously connected to the screened chameleon solution yields
\begin{align}
    Q_\mathcal{C}(r\ll R)
    \simeq
    &\frac{1}{2M_{\rm nl}^2}\!\Bigg[
      -M^2 \nonumber\\
      &+
      \sqrt{
        M^4
        +2M_{\rm nl}^2 (\alpha_B+2\alpha_M)\tilde{\rho}_m\delta
      }
    \Bigg].
    \label{eq:QCsmallr}
\end{align}

Importantly, this solution is independent of $r$, meaning the field is pinned to its effective minimum value within the high-density region, as is characteristic in the interior of a Chameleon-screened object.\footnote{Note that this relies on $M_{\rm nl}^2>0$. For theories predicting the opposite sign, the quantity inside the square-root in \eqref{eq:QCsmallr} would eventually become negative, mathematically invalidating this truncated approach. Alternatively, in such cases this equation would safely describe the field within a void, where $\delta<0$.} The shift of this potential minimum as a function of the local density $\delta$ is visualised by the solid markers in Fig.~\ref{fig:chameleon_potential_inverse}.

Far outside the object ($r > R$), the density approaches the background value and the effective mass is small, linearising the equation of motion to yield the general spherically symmetric Yukawa profile,\footnote{Note that regularity at infinity impedes the existence of $e^{+m_{\rm eff} r}$ solutions.}
\begin{align}
    Q_\mathcal{C}^{\rm(ext)}(r) = \frac{A}{r} e^{-m_{\rm eff} r},
    \label{eq:Q-cham-gen}
\end{align}
where $m_{\rm eff}^2 \equiv M^2/\Gamma$. Inside the bulk of the high-density object ($r < R - \Delta R$), the large nonlinear mass dynamically pins the field close to the minimum of its effective potential, meaning spatial gradients effectively vanish, $Q_\mathcal{C}' \simeq 0$.

Consequently, the exterior field is sourced almost entirely by a thin shell of thickness $\Delta R$ near the surface, where the field departs from its interior minimum and develops a non-negligible slope. Integrating the equation of motion \eqref{eq:QCeq} across this active shell region gives\footnote{Note that mass terms do not contribute significantly within the shell region compared to the driving kinetic term.}
\begin{align}
    \bigl[r^2 Q_\mathcal{C}'(r)\bigr]_{R-\Delta R}^{R} = -\frac{\alpha_B + 2\alpha_M}{2\Gamma}\int_{R-\Delta R}^{R} dr\, r^2 \tilde{\rho}_m\, \delta(r).
\end{align}
Assuming a constant density source for simplicity, the integral evaluates to the mass contained within the shell, $\mu_{\rm eff} \equiv \mu \frac{3\Delta R}{R}$. Since the interior derivative vanishes ($Q_\mathcal{C}'(R - \Delta R) \approx 0$), the surface flux evaluates strictly to
\begin{align}
    Q_\mathcal{C}'(R) \simeq -\frac{(\alpha_B + 2\alpha_M)\mu_{\rm eff}}{8\pi\Gamma R^2}.
    \label{eq:Qprime-cham}
\end{align}
Enforcing continuity of $Q_\mathcal{C}$ and $Q_\mathcal{C}'$ at the boundary $r=R$ fixes the amplitude $A$. Differentiating the exterior profile \eqref{eq:Q-cham-gen} and plugging it into the flux relation \eqref{eq:Qprime-cham} uniquely determines the exterior solution to be
\begin{align}
    Q_\mathcal{C}(r \gg R)
    &\simeq
    \frac{(\alpha_B + 2\alpha_M)\,\mu_{\rm eff}}
         {8\pi \Gamma\, r}\,
    \frac{e^{-m_{\rm eff}(r - R)}}{m_{\rm eff} R + 1}.
    \label{eq:thin-shell-sol}
\end{align}
For $3\Delta R / R \ll 1$ the object is strongly screened, while for $3\Delta R / R \ge 1$ the full mass sources the field and the screening disappears. 

It is important to emphasise that recovering this phenomenology here is highly non-trivial. Unlike standard Chameleon treatments that typically solve the full nonlinear potential exactly, our formulation employs a truncated perturbative framework. Within this approach, retaining perturbations up to second order successfully captures the leading-order physical signature of the mechanism: the density-dependent shift of the potential minimum and the resulting dynamical increase of the scalar field's effective mass in high-density environments. While extending the expansion to higher perturbative orders would theoretically increase the precision of the interior mass profile, this second-order treatment is \red{sufficient to capture the onset of the thin-shell effect}.

This physical behaviour is explicitly visualised in Figs.~\ref{fig:chameleon_Q_meff} and \ref{fig:chameleon}. Fig.~\ref{fig:chameleon_Q_meff} demonstrates how the effective mass $m_{\rm eff}^2$ is driven to higher constant values inside the dense core, dynamically locking the scalar field, while dropping sharply near the surface. Consequently, as shown in the upper panel of Fig.~\ref{fig:chameleon}, the resulting nonlinear scalar flux is heavily suppressed compared to the unscreened linear expectation, only beginning to grow near the surface where the thin shell activates. The analytical expression \eqref{eq:thin-shell-sol} is compared to the fully numerical solution in Fig.~\ref{fig:chameleon_Q_meff}, where we observe strong agreement with deviations occurring primarily in the complex thin-shell transition region.

\subsection{Phaedrus}\label{sec:phaedrus}
\begin{figure}
    \centering
    \includegraphics[width=\linewidth]{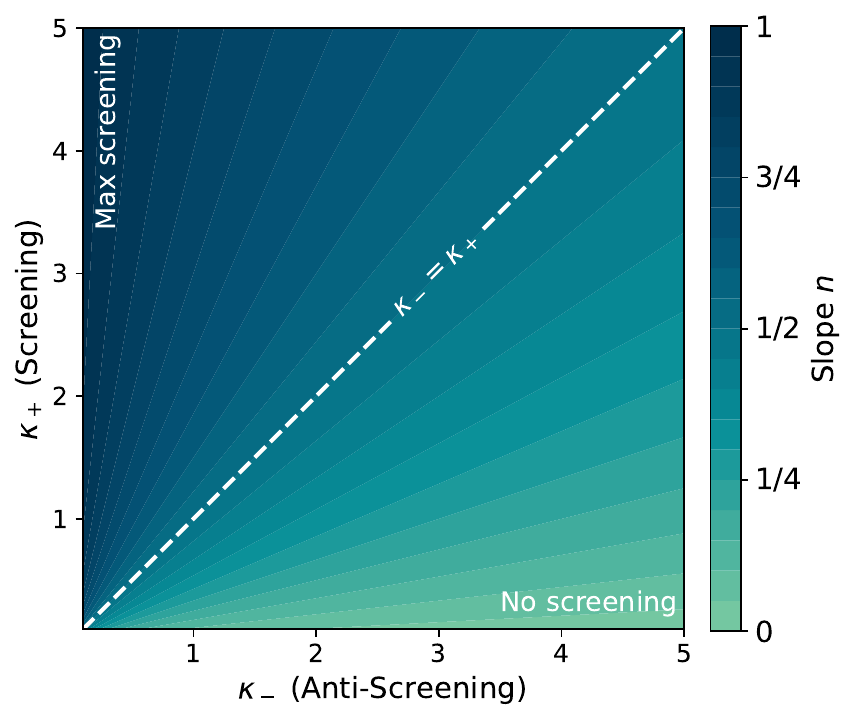}
    \caption{The screening efficiency $n$ of the Phaedrus varies with the ratio of pro-screening ($\kappa_{+}$) to anti-screening ($\kappa_{-}$) terms.}
    \label{fig:d2d3}
\end{figure}
\begin{figure}
    \centering
    \includegraphics[width=\linewidth]{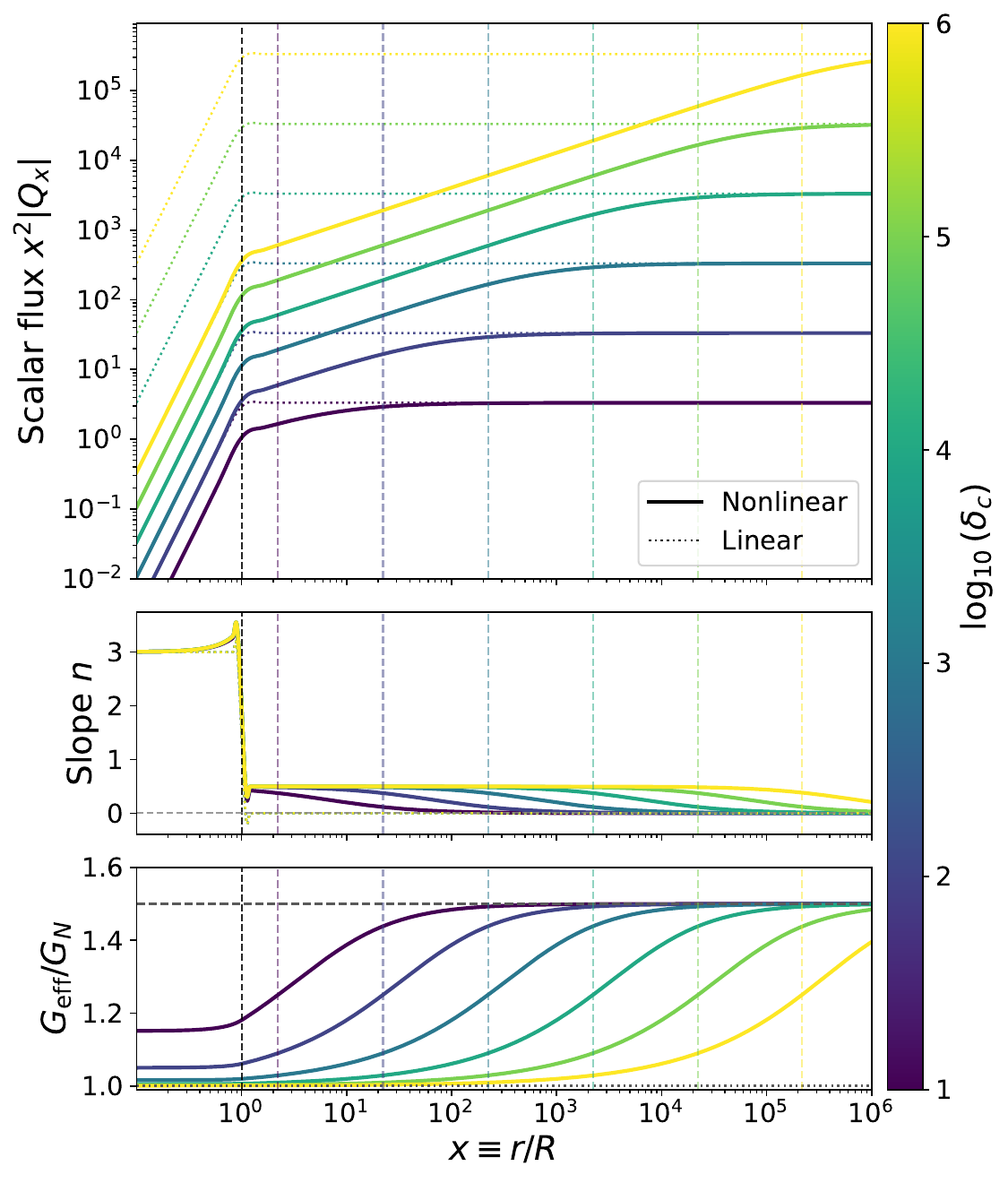}
    \caption{
    Numerical solutions of Phaedrus screening with $\kappa_{-}=\kappa_{+}$ as a function of the normalised radial coordinate for a spherical top-hat density (surface at $x=1$). 
    \textbf{Upper panel:} The scalar flux $x^{2}\lvert Q_x\rvert$. Solid curves represent the full nonlinear solutions, while dotted curves indicate the corresponding unscreened linear solutions. 
    \textbf{\red{Middle} panel:} The screening efficiency slope $n$~\eqref{eq:n_definition}, transitioning from the source interior ($n=3$ for $x < 1$) to the analytical Phaedrus plateau ($n=0.5$), before eventually decaying to the unscreened limit ($n=0$).
    \red{\textbf{Bottom panel}: the physical effective gravitational coupling $G_{\rm eff}/G_N=\Phi'/\Phi'_{\rm GR}$ (Eq.~\eqref{eq:Geff}). The dashed line marks the unscreened value $1+(\alpha_B+2\alpha_M)^2/2\Gamma$. Inside the screening radius, the flux is suppressed and $G_{\rm eff}/G_N$ returns to the GR value $1$ (dotted line).
    Different colours correspond to varying density amplitudes, demonstrating that denser sources linearly scale the Phaedrus radius $r_\mathcal{P}$ (marked for each solution by the corresponding vertical dashed lines).}
    }
    \label{fig:phaedrus_diagnostics}
\end{figure}
We now focus on the scenario where the operator $\mathcal{P}$ in Eq.~\eqref{eq:scalar-pert-eom2} provides the dominant nonlinear contribution. This represents a novel \red{candidate} screening phenomenology introduced in this paper, which we term the \textit{Phaedrus} mechanism.\footnote{The name is inspired by the chariot allegory in Plato's \textit{Phaedrus}. Much like the charioteer's two horses pulling in opposing directions (one toward the divine and one toward the earth) the efficiency of this screening mechanism is determined by the tension between the pro-screening $\kappa_{+}$ and anti-screening $\kappa_{-}$ interactions.} The Phaedrus operates by suppressing the fifth force through field-dependent non-canonical kinetic interactions: $Q\nabla^2 Q$ and $(\nabla Q)^2$.
\red{An explicit luminal Horndeski model realising the Phaedrus regime is readily constructed with a field-dependent non-canonical kinetic term, $K(\phi,X)=X+g(\phi)X^2/\Lambda^4$, combined with a non-minimal coupling $G_4(\phi)$. This yields non-zero $\kappa_\pm$ through $K_{XX}$ and $K_{\phi X}$ (see Eq.\eqref{eq:d2d3-reduced-K}), while switching off the Vainshtein operator (no $G_3$) and retaining a matter source.\footnote{\red{The nonlinear $X$-dependence is essential here, as a mere field-dependent normalisation $\propto f(\phi)X$ can be removed by a field redefinition and sources no physical screening.}}}

A defining feature is that the screening radius scales linearly with the source mass ($r_\mathcal{P} \propto \mu$), implying that the screened volume per unit mass grows as $\mu^2$. Consequently, massive objects such as galaxy clusters develop proportionally larger screened envelopes compared to their constituent galaxies, providing a distinct observational signature. However, as will be detailed below, the theoretical viability of this extended screening regime relies on highly non-trivial theoretical conditions. \red{We therefore regard Phaedrus as a conjectural mechanism, pending a full time-dependent stability analysis.}

Let us first understand how this term screens fifth forces in practice, before contextualising the theoretical setups in which it can be activated. The master equation for the scalar perturbation $Q$ takes the form
\begin{align}
\label{eq:Phaedrus}
\Gamma(r^2Q_\mathcal{P}')' &+ \kappa_{-} Q (r^2 Q_\mathcal{P}')' + \kappa_{+} r^2 (Q_\mathcal{P}')^2 \nonumber\\
&= -(\alpha_B+2\alpha_M)r^2\tilde{\rho}_m\delta,
\end{align}
where recall that in Eq.~\eqref{eq:relabel-coeffs} we defined $\kappa_{-}$ and $\kappa_{+}$ respectively as the coefficients $D_6^{QQ}$ \eqref{eq:D6QQ} and $D_7^{QQ}$ \eqref{eq:D7QQ}.

\subsubsection{General Screening Behaviour ($\kappa_{-} \neq \kappa_{+} \neq 0$)}
Outside the source ($r>R$), the density contrast vanishes, and the profile is determined by the nonlinear terms. We seek a power-law solution of the form $Q(r) \propto r^{n-1}$, where $n$ is the screening efficiency~\eqref{eq:n_definition}. Substituting this ansatz into the exterior equation ($\kappa_{-} Q_\mathcal{P} (r^2 Q_\mathcal{P}')' + \kappa_{+} r^2 (Q_\mathcal{P}')^2 = 0$), we find that both terms scale identically as $r^{2n-2}$. The characteristic equation for the slope $n$ then becomes
\begin{equation}
(n-1) [ n \kappa_{-} + (n-1) \kappa_{+} ] = 0.
\end{equation}
Discarding the trivial root $n=1$,\footnote{The root $n=1$ yields a constant field profile ($Q \propto r^0$), corresponding to a trivial solution with identically vanishing spatial gradients ($Q'=0$), meaning no fifth force is generated.} we find the physical root to be
\begin{equation}
n = \frac{\kappa_{+}}{\kappa_{-} + \kappa_{+}}.
\end{equation}
This result, visualised in Fig.~\ref{fig:d2d3}, highlights that the screening efficiency is non-universal and depends on the competition between the `anti-screening' term $\kappa_{-}$ and the `pro-screening' term $\kappa_{+}$. In other words, the profile interpolates between a Newtonian-like regime ($n \to 0$) when $\kappa_{-} \gg \kappa_{+}$ and a strongly screened regime ($n \to 1$) when $\kappa_{+} \gg \kappa_{-}$.

\blue{A natural question is whether this classification is invariant under a redefinition of the scalar. The effective coupling $G_{\rm eff}/G_N$ of Eq.~\eqref{eq:Geff} is invariant by construction, being built from the metric potential; we now argue that so is $n$. First, a change of the perturbation variable, e.g. among the conventions used here ($Q$, $v_X$ and $\delta\phi$; Appendix~\ref{app:conversions}), merely rescales $Q$ by a spatially-uniform, time-dependent factor. Being independent of $r$, this factor cancels in the radial derivative $n=\mathrm{d}\ln|r^2 Q'|/\mathrm{d}\ln r$, so $n$ is identical in all three conventions, even though the individual coefficients $\kappa_{-}$ and $\kappa_{+}$ do rescale, and are thus convention-dependent, exactly as the effective mass is (Appendix~\ref{app:conversions}). Second, one may ask whether the screening itself survives a general field redefinition $\phi=f(\chi)$, which reshuffles the coefficients $\kappa_{\pm}$. It does: the fifth force is the deviation of the metric potential from its GR value (see Eq.~\eqref{eq:fifth-force}), and redefining the scalar leaves the spacetime metric untouched, so the suppressed force profile is necessarily invariant, and with it the slope $n$ and the radius $r_\mathcal{P}\propto\mu$. This is exactly the content of the frame-independent coupling $G_{\rm eff}/G_N$, whose return to unity inside $r_\mathcal{P}$ (bottom panel of Fig.~\ref{fig:phaedrus_diagnostics}) is a statement about the physical potential rather than the scalar variable. Thus, although a redefinition shuffles $\kappa_{+}$ and $\kappa_{-}$ separately, the competition between them that fixes $n$ is pinned by this invariant profile, and so cannot be redefined away.}

The full screened solution can be approximated by matching this power law to the linear solution $Q_{\rm lin}$ in Eq.~\eqref{eq:Q_Newtonian} at a characteristic screening radius $r_\mathcal{P}$:
\begin{equation}
Q_\mathcal{P}(r) \approx Q_{\rm lin}(r_\mathcal{P}) \left( \frac{r}{r_\mathcal{P}} \right)^{n-1}, \quad \text{for } r \ll r_\mathcal{P}.
\end{equation}
By equating the magnitude of the linear and nonlinear terms at the transition, we find that the screening radius scales linearly with the source mass, $r_\mathcal{P} \propto \mu$.\footnote{In the unscreened exterior, the linear kinetic term scales as $\nabla^2 Q \sim \mu/r^3$. The Phaedrus mechanism is driven by the interactions $Q \nabla^2 Q$ and $(\nabla Q)^2$, both of which scale as $\mu^2/r^4$. Equating the linear and nonlinear terms at the screening boundary then yields the linear relationship $r_\mathcal{P} \propto \mu$.} \red{We emphasise that this is a scaling argument, valid for general $\kappa_-\neq\kappa_+$. A closed-form profile, and hence the exact prefactor of $r_\mathcal{P}$, is only obtainable in the special cases treated analytically below.}

For standard Vainshtein screening, $r_\mathcal{V} \propto \mu^{1/3}$ implies that the volume of space screened per unit mass remains constant regardless of the source mass. For $K$-mouflage we typically have that $r_\mathcal{K} \propto \mu^{1/2}$~\cite{Brax:2014wla}, meaning the screened volume per mass grows weakly with the mass. Finally, Phaedrus yields a screened volume that scales as the cube of the mass ($V \propto r_\mathcal{P}^3 \propto \mu^3$), meaning the screened volume per unit mass grows as $\sim\mu^2$. Consequently, highly massive structures could exhibit vast screened envelopes, providing a pronounced and unique phenomenological signature.
\red{Additionally, note that in the bottom panel of Fig.~\ref{fig:phaedrus_diagnostics}, we show that the suppression of the scalar flux propagates to the metric potential, driving $G_{\rm eff}/G_N\to1$ inside the screening radius (see Eq.~\eqref{eq:Geff}), and therefore demonstrating that the recovery of GR holds at the level of the dynamical metric potential and not merely for the scalar profile.}

\begin{figure}
    \centering
    \includegraphics[width=\linewidth]{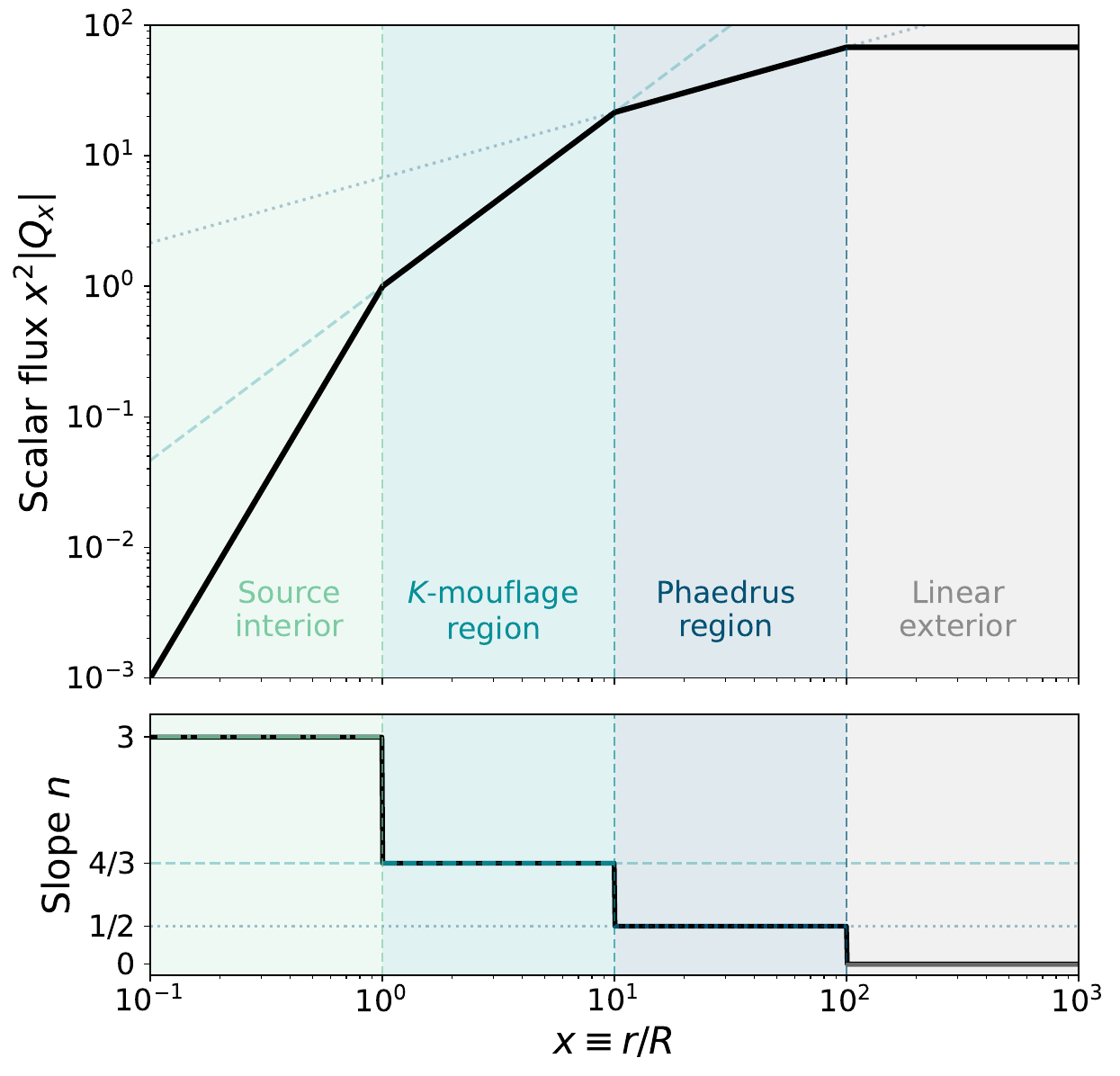}
    \caption{
    Toy example of the hierarchical `shell' structure of screening around a massive source when both $K$-mouflage and Phaedrus interactions are active. 
    \textbf{Upper panel:} The scalar flux $x^{2}\lvert Q_x\rvert$. \textbf{Lower panel:} The local screening efficiency $n$~\eqref{eq:n_definition} 
    As spatial gradients steepen towards the source ($x \to 0$), the $K$-mouflage operator dominates the inner halo ($n=4/3$). As the gradients weaken at larger radii, the field dynamically transitions into the Phaedrus regime ($n=1/2$) before eventually yielding to the unscreened linear profile ($n=0$).
    }
    \label{fig:hierarchical_shells}
\end{figure}

\subsubsection{Exact Analytical Solution ($\kappa_{-} = 0$)}
In the limit where the anti-screening operator vanishes ($\kappa_{-} = 0$), the exterior equation of motion reduces to a balance between the linear kinetic term and the pure gradient-squared interaction, $\Gamma(r^2Q_\mathcal{P}')' + \kappa_{+} r^2 (Q_\mathcal{P}')^2 = 0$. By casting this as a first-order Bernoulli differential equation for $Q_\mathcal{P}'$, one can obtain an exact algebraic solution which scales as $Q_\mathcal{P}' \propto 1/r$ in the nonlinear regime, meaning the scalar field adopts a logarithmic profile, $Q_\mathcal{P}(r) \propto \ln(r)$. This corroborates the limit of our power-law ansatz ($n \to 1$), confirming that the maximal screening limit corresponds to a smooth transition from a power-law suppression into a logarithmic one.

\subsubsection{Exact Analytical Solution ($\kappa_{-} = \kappa_{+}$)}
For the specific symmetric case where $\kappa_{-} = \kappa_{+} \equiv \kappa$ (and hence $n=1/2$), Eq.~\eqref{eq:Phaedrus} allows for a first and second integral, resulting in
\begin{equation}
\frac{\kappa}{2} Q_\mathcal{P}^2 + \Gamma Q_\mathcal{P} = \Phi_N(r) + C,
\end{equation}
where $\Phi_N(r)$ is the effective Newtonian potential generated by the source mass defined as
\begin{equation}
\Phi_N(r) \equiv -(\alpha_B+2\alpha_M) \int^r \frac{\mu(\tilde{r})}{\tilde{r}^2} d\tilde{r}.
\end{equation}
Solving for $Q_\mathcal{P}$, we find the exact profile:
\begin{equation}
Q_\mathcal{P}(r) = \frac{-\Gamma + \sqrt{\Gamma^2 + 2 \kappa \Phi_N(r)}}{\kappa},
\end{equation}
where we have set the integration constant $C=0$ to satisfy the boundary condition $Q_\mathcal{P}(\infty)=0$.
Explicitly, outside a source of mass $\mu$, this solution is:
\begin{equation}
Q_\mathcal{P}^{(out)}(r) = \frac{-\Gamma}{\kappa} \left[ 1 - \sqrt{1 + \frac{2 \kappa (\alpha_B+2\alpha_M)\mu}{\Gamma^2 r}} \right],
\end{equation}
which confirms the general scaling arguments: at large distances ($r \to \infty$), we recover the linear Newtonian limit, while at short distances, the term under the square root dominates, leading to a $r^{-1/2}$ decay ($n=1/2$). The screening radius is readily identified from the square root term as
\begin{align}
    r_\mathcal{P} =\frac{2 \kappa (\alpha_B+2\alpha_M)\mu}{\Gamma^2}.
\end{align}

As briefly noted above, the linear dependence of the screening radius on the source mass carries distinct observational implications that cleanly separate Phaedrus from other established mechanisms.

\subsubsection{Physical viability}
\label{sec:phaedrus-viability}
To contextualise the physical viability of the Phaedrus regime, we must examine its competition with both the standard linear kinetic term and higher-order kinetic nonlinearities. To isolate this effect, we restrict our focus to theories governed by the reduced Lagrangian $\mathcal{L}_\phi=G_4(\phi)R+K(\phi,X)$, \blue{which removes the cubic Galileon and hence the Vainshtein operator. Field dependence in $K$ and $G_4$ can still source a Chameleon-type effective mass; here we assume this effective mass to be negligible, neglecting the corresponding $M_{\rm nl}^2 Q^2$ contribution, in order to isolate the competition between the kinetic Phaedrus and $K$-mouflage operators.} Evaluating the exact analytical coefficients reveals that Phaedrus is sourced by non-canonical kinetic terms ($K_{XX}, K_{XXX}$) and shift-symmetry breaking interactions ($K_{\phi X}$, $K_{\phi XX}$), coupled to the cosmological background ($\dot{\phi}$, $\ddot{\phi}$, and $H$), see Eqs.~\eqref{eq:d2d3-reduced-K}. \red{A concrete example is the minimal model introduced above, $K(\phi,X)=X+g(\phi)X^2/\Lambda^4$ with a non-minimal coupling $G_4(\phi)$, for which these coefficients are non-zero.} Extending to third order in perturbations to capture the leading $K$-mouflage contribution, the equation of motion for the scalar perturbation takes the form:\footnote{This equation has been obtained using \href{https://github.com/sergisl/xAlpha}{\faGithub \;xAlpha}. Note as well that we again use $Q$ instead of $Q_{\mathcal{P}}$, as solutions associated with this equation are not purely of Phaedrus type but rather contain mixed screening.}
\begin{align}
\label{eq:cubicQ}
&\Gamma(r^2Q')' + \underbrace{\kappa_{-} Q (r^2 Q')' + \kappa_{+} r^2 (Q')^2}_{\text{Phaedrus}}\nonumber\\
&\qquad + \underbrace{K_{XX}(Q')^2(r^2Q')'}_{K\text{-mouflage}} = -(\alpha_B+2\alpha_M)r^2\tilde{\rho}_m\delta.
\end{align}

\paragraph{Competition with the Linear Term:}
For Phaedrus to dominate the outer regions, the second-order terms must surpass the linear kinetic term, roughly requiring $Q \gtrsim \Gamma / \kappa_\pm$. This can be achieved in two distinct ways. The first is if the linear kinetic term is inherently suppressed ($\Gamma \ll 1$). A finite but suppressed $\Gamma$ ensures Phaedrus operates over vast scales while safely yielding to the standard linear regime in the deep cosmological background, preventing unbounded fifth forces. We note that the limit $\Gamma \ll 1$ raises important caveats regarding the QSA and dynamical stability. Because the effective spatial kinetic term is strictly proportional to the field's sound speed ($\Gamma \equiv D c_s^2$), suppressing it (assuming an order-unity $D$) drives $c_s^2 \to 0$. This limit inherently breaks the QSA and has been shown to potentially trigger dynamical instabilities \cite{Hassani:2021tdd,Hassani:2022xyb,Eckmann:2022wtd}. Resolving the ultimate stability of these models therefore requires a full time-dependent analysis, which is left for future work. Finally, note that this suppressed kinetic limit is satisfied by specific theoretical setups, such as Cuscuton-like models \cite{Afshordi:2006ad,Iyonaga:2018vnu} or theories near a Ghost Condensate limit \cite{ArkaniHamed:2003uy}.\footnote{It is worth noting that a pure Cuscuton theory has an infinite speed of sound and lacks propagating scalar degrees of freedom on the background, meaning it does not typically mediate a standard fifth force requiring screening.}Alternatively, rather than suppressing $\Gamma$, the required hierarchy can be achieved if the nonlinear coefficients are enhanced ($\kappa_\pm \gg \Gamma$). Analogous to Chameleon-type models, this enhancement can arise naturally from nonlinearities in $K(\phi, X)$. While this avoids the vanishing sound speed issues associated with $\Gamma \ll 1$, it implies, as discussed in the introduction, that the expansion around the cosmological background will eventually break down deep inside the screened region. Nevertheless, while the series cannot be safely truncated infinitely deep inside a dense source, the emergence of this hierarchy confirms that our framework successfully captures the leading-order physical onset of the Phaedrus mechanism.\footnote{We thank Kazuya Koyama for related useful insights.}

\paragraph{Competition with $K$-mouflage:}
Near a dense object ($r \to 0$), spatial gradients steepen much faster than the field amplitude. Comparing the operators in Eq.~\eqref{eq:cubicQ}, the cubic $K$-mouflage term ($\sim (Q')^3/r$) will inevitably overcome the quadratic Phaedrus term ($\sim Q Q'/r^2$) at small radii. Consequently, in theories permitting $K$-mouflage, Phaedrus cannot exist as the deepest interior mechanism. Instead, it manifests as an \emph{intermediate shell}: a transitional screening regime seamlessly sandwiched between the linear Newtonian profile at large radii and the deeply nonlinear $K$-mouflage core (see Fig.~\ref{fig:hierarchical_shells} for a toy visualisation).\footnote{Note that similar hierarchical `shell' structures have already been encountered regarding higher-order Galileon terms (absent in this work due to the luminality requirement) \cite{Burrage:2010rs}.}

\paragraph{Observational Signatures:}
This hierarchical `shell' structure carries distinct observational implications. Because the Phaedrus screening radius scales linearly with the source mass ($r_\mathcal{P} \propto \mu$), it grows much faster than a standard dark matter halo ($R_{\rm vir} \propto \mu^{1/3}$). For massive galaxy clusters, the ratio $r_\mathcal{P}/R_{\rm vir}$ becomes large, pushing the Phaedrus boundary deep into the surrounding cosmic web. \red{We speculate that this could create} a unique spatial signature, in which inner cluster dynamics are governed by standard $K$-mouflage while the extended outskirts feel the $\sim r^{-1/2}$ Phaedrus suppression. To isolate this signature, one must probe the extreme outer regions of halos. Two ideal testbeds are the splashback radius \cite{Adhikari:2014lna} (sensitive to the exact gravitational force law in the accretion zone) and wide-field weak lensing shear maps \cite{Umetsu:2020wlf}, which can capture the extended spatial envelope of the effective gravitational potential.

\section{Conclusions}\label{sec:conclusions}
Testing gravity with the latest and upcoming cosmological surveys requires a robust understanding of gravity theories in the nonlinear regime. In particular, determining which screening mechanisms are activated by a given theory, and to what extent, is crucial to systematically explore modifications of gravity on these scales.
In this paper, we have constructed a unified framework to simultaneously capture diverse screening mechanisms within luminal Horndeski theories. Our key findings are as follows:
\begin{itemize}
    \item We have derived and organised, for the first time, the complete set of second-order cosmological perturbation equations for general luminal Horndeski theories without the use of the quasistatic and weak-field approximations. These are shown in Eqs. \eqref{eq:matrix-eoms} to \eqref{eq:S2}, with the coefficients written in the $\alpha$-basis in Appendix~\ref{app:expressions}.
    \item Employing the aforementioned approximations, we have shown that surviving nonlinear corrections are confined to the scalar field equation, while the metric equations remain well-described by linear theory.
    \item By identifying the resulting scalar nonlinear operators as respective sources for different screening mechanisms, we have derived a master screening equation for luminal Horndeski theories \red{to second order in perturbations}.
    We have demonstrated how this equation, truncated at second order, simultaneously recovers the established Vainshtein mechanism \red{and the onset of Chameleon screening} directly from the covariant theory. This is highly non-trivial, especially for the Chameleon mechanism, which is usually studied within distinctly different frameworks.
    \item We have also identified a novel \red{candidate} kinetic screening regime, termed Phaedrus screening, sourced by non-canonical kinetic terms. While the radial suppression of the fifth force within this regime is milder than in the standard Vainshtein or $K$-mouflage mechanisms, it exhibits a distinctively extended screening envelope: its screening radius scales linearly with the source mass ($r_\mathcal{P} \propto \mu$). This scaling is significantly steeper than in other kinetic mechanisms, making its footprint amplified and potentially dominant for massive structures. Consequently, \red{if physically realised}, its effects could yield novel observable signatures specifically in the outskirts of galaxy clusters. \blue{However, its activation requires a non-trivial hierarchy in which the quadratic operators dominate the linear kinetic term, achieved either through a strongly suppressed linear kinetic term ($\Gamma \ll 1$) or through enhanced nonlinear coefficients ($\kappa_\pm \gg \Gamma$). As discussed in Sec.~\ref{sec:phaedrus-viability}, each route carries its own caveat, respectively regarding the nonlinear dynamical stability of the system and the validity of the perturbative truncation deep inside the source.}
    \item We have introduced two open-source software packages developed for this work: \href{https://github.com/sergisl/xAlpha}{\faGithub \;xAlpha}, a \texttt{Mathematica} suite capable of automatically deriving nonlinear perturbation equations in luminal Horndeski theories and extracting the relevant coefficients, and \href{https://github.com/sergisl/escut}{\faGithub \;escut}, a dedicated \texttt{Python} module designed to numerically integrate the master screening equation. These are intended to facilitate future research and allow for the independent reproduction of our results.
\end{itemize}
This work opens several avenues for future research:
\begin{itemize}
    \item \textbf{Extension to third-order perturbations:} Extending the perturbative framework to third order is essential to capture mechanisms driven by higher-order terms, such as $K$-mouflage and the Symmetron. Furthermore, this would allow for a more precise assessment of the Chameleon, determining the specific impact of higher-order corrections on the leading-order results presented here. Finally, the systematic exploration of third-order perturbations could reveal new screening operators.
    \item \textbf{Simultaneous screening mechanisms:} The master screening equation can be numerically solved to investigate the simultaneous interplay between different screening mechanisms. While the dedicated numerical solver developed for this framework, \href{https://github.com/sergisl/escut}{\faGithub \;escut}, has been robustly validated for isolated screening regimes, simultaneously ensuring the strict resolvability of each individual operator introduces distinct challenges that require further investigation.
    \item \textbf{Assessing the physical viability of the Phaedrus regime:} In this work, we have identified that quadratic spatial operators, such as $(\nabla Q)^2$ and $Q \nabla^2 Q$, offer a novel route to effectively suppress fifth forces. However, achieving the necessary hierarchy for this mechanism to dominate introduces distinct physical and mathematical challenges. If this hierarchy is achieved via a suppressed linear kinetic term ($\Gamma \ll 1$), it risks breaking the quasi-static approximation and triggering dynamical instabilities. Alternatively, achieving it via enhanced nonlinear coefficients ($\kappa_\pm \gg \Gamma$) avoids these issues, but formally challenges the validity of the perturbative expansion deep inside the source. Nevertheless, much like the Chameleon mechanism, this breakdown might indicate that our framework is capturing the effective limit of an exact non-perturbative Phaedrus mechanism. Determining the ultimate physical fate of these models, whether by resolving the time-dependent stability of the former regime, or by formally constructing the exact non-perturbative completion of the latter, represents a critical direction for future investigation.
    \item \textbf{Implementation in \texttt{Hi-COLA} and Stage IV observables:} To effectively confront modified gravity theories with data, these derived screening profiles must be integrated into fast simulation techniques such as \texttt{Hi-COLA}. This integration is essential for producing approximate N-body simulations, from which one can then extract the robust summary statistics necessary to test gravity against nonlinear observables.
\end{itemize}

\section*{Acknowledgments}
We thank Emilio Bellini, Clare Burrage, Kazuya Koyama, Daniela Saadeh and Obinna Umeh for useful discussions.
SS, TB and KN are supported by ERC Starting Grant SHADE (grant no. StG 949572). TB is further supported
by a Royal Society University Research Fellowship (grant no.URF\textbackslash R\textbackslash 231006). JH is supported by a PhD studentship from UKRI-STFC. 
In deriving the results of this paper, we have used \href{https://github.com/sergisl/escut}{\faGithub \;escut}~\cite{escut} and \href{https://github.com/sergisl/xAlpha}{\faGithub \;xAlpha}~\cite{xAlpha}, the latter being a package based on \texttt{xAct} \cite{xAct} and \texttt{xPand}~\cite{Pitrou:2013hga}.
For the purpose of open access, the authors have applied a Creative Commons Attribution (CC BY) licence to any Author Accepted Manuscript version arising from this work.
\appendix

\section{Covariant equations and background stability}
\label{app:cov-eoms}
Here, we provide the fully covariant equations of motion for luminal Horndeski theories as described by the action~\eqref{eq:S2}.\footnote{These can be easily obtained from the equations in \cite{Kobayashi:2025evr} for generic cubic higher-order scalar-tensor (HOST) theories, and have in fact been derived using the resources developed there.} First, the metric EOMs are given by
\begin{align}
& -2G_4 G_{\mu\nu} +\phi_\mu\phi_\nu\left(K_X-2G_{3\phi}+2G_{4\phi\phi}-G_{3X}\Box\phi\right)\nonumber\\
& + g_{\mu\nu}\big[K-2X(G_{3\phi}-2G_{4\phi\phi})+G_{3X}\phi^\alpha\phi^\beta\phi_{\alpha\beta}\nonumber\\
&-2G_{4\phi}\Box\phi\big]+ 2G_{4\phi}\phi_{\mu\nu} + 2G_{3X}\phi_\mu\phi_{\nu\sigma}\phi^\sigma\nonumber \\
&   = -(\rho + p) u_\mu u_\nu - p g_{\mu\nu}.
\end{align}
Second, the scalar EOM is given by
\begin{align}
& K_\phi + G_{4\phi} R -2X(K_{\phi X}-G_{3\phi\phi})+ \Box\phi\big[K_X\nonumber\\
&-2G_{3\phi}+2XG_{3\phi X}-G_{3X}\Box\phi+G_{3XX}\phi^\mu\phi^\nu\phi_{\mu\nu}\big] \nonumber\\
& +G_{3X}\left[R_{\mu\nu}\phi^\mu\phi^\nu+\phi_{\mu\nu}\phi^{\mu\nu}\right] -G_{3XX}\phi^\mu\phi^\nu\phi_\mu^\sigma\phi_{\sigma\nu}\nonumber\\
&-\phi^\mu\phi^\nu\phi_{\mu\nu}(K_{XX}-2G_{3\phi X})=0.
\end{align}
When substituting the background metric and scalar~\eqref{eq:bg}, these become the expressions shown in the main text~\eqref{eq:bg-eqs}. To ensure the background is stable, the following conditions need to be satisfied. First,
\begin{align}
        Q_S = \frac{2M_*^2 D}{(2-\alpha_B)^2} > 0, \qquad D = \alpha_K + \frac{3}{2}\alpha_B^2
\end{align}
ensures no ghost instabilities. Second, in order to not have gradient instabilities, we require
\begin{align}
        c_S^2 = \frac{1}{D} \left[\gamma_B - \gamma_E - 2(\alpha_B-\alpha_M)-\frac{1}{2}\alpha_B(\alpha_B+4\alpha_M)\right] > 0.
\end{align}
Note that a third type, namely tachyonic instabilities, can arise when the effective mass squared is negative, although these can often be rendered harmless if their growth rate is slower than the Hubble expansion.

\section{Dictionary of scalar perturbations}\label{app:conversions}
Different conventions exist for defining scalar field perturbations. In this Appendix, we show explicitly how the perturbation equations change accordingly. The perturbed scalar field is initially defined as
\begin{align}
\phi(t, \mathbf{x}) &= \phi(t) + \delta\phi(t, \mathbf{x}),
\end{align}
where $\phi(t)$ is the background scalar and $\delta\phi$ is its perturbation. One can simply use this as the perturbation field, as is done in \cite{DeFelice:2011hq}. In this paper, we have however followed \cite{Kimura:2011dc,Wright:2022krq} and redefined it as
\begin{eqnarray}
Q\equiv H\frac{\delta\phi}{\dot\phi}=\frac{\delta\phi}{d\phi/d\ln a}.
\end{eqnarray}

Alternatively, one can also employ the following redefinition, used in \cite{Bellini:2014fua},
\begin{eqnarray}
v_X\equiv -\frac{\delta\phi}{\dot\phi}.
\end{eqnarray}

Unsurprisingly, the form of the coefficients in the perturbation equations will change depending on which variable is used, hindering slightly their direct comparison. For instance, what we define as the effective mass of the scalar perturbation will necessarily differ for the different definitions. Nonetheless, it has been shown that limits on both large and small scales coincide for different definitions in terms of their prediction for $\mu$ and $\Sigma$ parameters~\cite{Pace:2020qpj}. Let us show here explicitly, as a simple example, how we can check the equivalence of the coefficients for terms in the linear $(00)$ metric equation. We can convert from $Q$ to $v_X$ and $\delta\phi$ with\footnote{Similarly, one can convert from $v_X$ and $\delta\phi$ to $Q$ by reversing the chain rule.}
\begin{align}
    Q&=H\frac{\delta\phi}{\dot\phi}=-Hv_X,\\
    \dot Q&=\frac{H}{\dot\phi}\left(\frac{\dot H}{H}-\frac{\ddot\phi}{\dot\phi}\right)\delta\phi+H\frac{\dot{\delta\phi}}{\dot\phi}\\
    &=-\dot Hv_X-H\dot v_X,\\
    \ddot Q&=\frac{1}{\dot\phi}\Bigg[\left(\ddot H-\frac{2\dot H\ddot\phi}{\dot\phi}-\frac{H\dddot\phi}{\dot\phi}+\frac{2H\ddot\phi^2}{\dot\phi^2}\right)\delta\phi\nonumber\\
    &\qquad+2\left(\dot H-\frac{H\ddot\phi}{\dot\phi}\right)\dot{\delta\phi}+H\ddot{\delta\phi}\Bigg]\\
    &=-\ddot Hv_X-2\dot H\dot v_X-H\ddot v_X,\\
    \nabla^2Q&=\frac{H}{\dot\phi}\nabla^2\delta\phi=-H\nabla^2v_X.
\end{align}

Using these relations, we can rewrite the scalar field perturbation terms in the linear $(00)$ metric equation as
\begin{align}
    {\cal \tilde{E}}^{(1)}&\ni H^2A_1^QQ+HA_2^Q\dot Q-\frac{1}{a^2}A_3^Q\nabla^2Q\\
    &=-3\left[H(2\dot H+\tilde{\rho}_m+\tilde{p}_m)-\dot H\alpha_B\right]Hv_X\nonumber\\
    &\qquad-(\alpha_K+3\alpha_B)H^2\dot v_X+\frac{1}{a^2}\alpha_BH\nabla^2v_X,\label{eq:comp-EB}\\
    &=\frac{1}{M_*^2}\Bigg[\mu\delta\phi+\frac{2}{\dot\phi}(\Sigma+3H\Theta)\dot{\delta\phi}\nonumber\\
    &\qquad+\frac{1}{a^2}\frac{2}{\dot\phi}(\Theta-H\mathcal{G}_T)\nabla^2\delta\phi\Bigg],\label{eq:comp-dFK}
\end{align}
where we have substituted the relevant expressions for the coefficients in Appendix~\ref{app:expressions}. On one hand, Eq.~\eqref{eq:comp-EB} is expressed in terms of $v_X$, and matches the corresponding terms in Eq.~(3.17) in \cite{Bellini:2014fua}. On the other hand, Eq.~\eqref{eq:comp-dFK}, which matches the result in Eq. (26) in \cite{DeFelice:2011hq}, is expressed in terms of $\delta\phi$ and the functions $\mu$, $\Sigma$, $\Theta$ and $\mathcal{G}_T$, whose definition can be found in \cite{DeFelice:2011hq}.

Having understood how the mappings between different definitions of the perturbed scalar work, one can similarly convert the rest of the perturbation equations. Nonetheless, let us show here explicitly the conversion of the linear effective mass of the scalar field. Using the variable $Q$, we define the mass term\footnote{Note that here we renamed $M^2\equiv M_Q^2$ in order to compare this to the mass terms in other conventions, i.e. $M^2_{\delta\phi}$ and $M^2_{v_X}$.}
\begin{align}
    M^2&\equiv M^2_Q\equiv -H^2D_1^Q\nonumber\\
    &=-\Bigg[\dot H(-3\gamma_E+3\gamma_B+\gamma_K)+\frac{\ddot H}{H}(\alpha_K+3\alpha_B)\Bigg].
\end{align}

Using $v_X$ (and the corresponding chain rule transformations), the effective mass term for scalar perturbations becomes
\begin{align}
    M^2_{v_X}&\equiv -H^2D_1^Q-\dot HD_2^Q-\frac{\ddot H}{H}D_3^Q\nonumber\\
    &=-3\left[\dot H(-\gamma_E+\gamma_B)+\frac{\ddot H}{H}\alpha_B\right],
    \label{eq:M2vX}
\end{align}
which matches the expression in Eq. (3.22) in~\cite{Bellini:2014fua}. Note, however, that in comparison with their expression, we have expressed $M_{v_X}^2$ (called $M^2$ in~\cite{Bellini:2014fua}) in terms of the newly defined $\gamma$ functions. We can then see that the difference with respect to $M^2_Q$ is
\begin{align}
    M^2_{v_X}=M^2_Q+\dot H\gamma_K+\frac{\ddot H}{H}\alpha_K,
\end{align}
where the difference arises from time derivative terms being converted into mass terms, i.e. $\dot Q\rightarrow v_X$, or vice versa. Note that these are precisely the $D_2^Q$ and $D_3^Q$ terms~\eqref{eq:D23Q}.

Using $\delta\phi$ (and the corresponding chain rule transformations), the effective mass term for scalar perturbations becomes
\begin{align}
    M^2_{\delta\phi}&\equiv -\frac{H}{\dot\phi^2}\Bigg[H^3D_1^Q+H^2\left(\frac{\dot H}{H}-\frac{\ddot\phi}{\dot\phi}\right)D_2^Q\nonumber\\
    &\qquad+\left(\ddot H-\frac{2\dot H\ddot \phi}{\dot\phi}-\frac{H\dddot\phi}{\dot\phi}+\frac{2H\ddot\phi^2}{\dot\phi^2}\right)D_3^Q\Bigg]\nonumber\\
    &=\frac{1}{\dot\phi}\left[\dot\mu+3H(\mu+\nu)\right],
    \label{eq:M2dphi}
\end{align}
which matches the expression in Eq. (3.22) in~\cite{DeFelice:2011hq}. As discussed in the main text, for strictly shift-symmetric theories (where a standard bare potential mass $V_{\phi\phi}$ is forbidden), this effective mass $M^2_{\delta\phi}$ exactly vanishes. This is because the second-order action for $\delta\phi$ must remain invariant under constant shifts $\delta\phi \to \delta\phi + c$, explicitly prohibiting a $\delta\phi^2$ term. It is only when switching to the variable $Q$ (or equivalently $v_X$) that a shift-symmetric effective mass term $M^2_Q$ is generated, arising from the time evolution of the background metric. Finally, we can see that the difference with respect to $M^2_{v_X}$ is
\begin{align}
    M_{\delta\phi}^2&=\frac{H}{\dot\phi^2}\Bigg[M_{v_X}^2-\frac{H^2\ddot\phi}{\dot\phi}\gamma_K\nonumber\\
    &\qquad+\frac{1}{\dot\phi}\left(2\dot H\ddot\phi+H\dddot\phi-\frac{2H\ddot\phi^2}{\dot\phi}\right)\alpha_K\Bigg].
\end{align}

\section{Numerical solver and convergence tests}\label{app:convergence}
\begin{figure*}
    \centering
    \includegraphics[width=\linewidth]{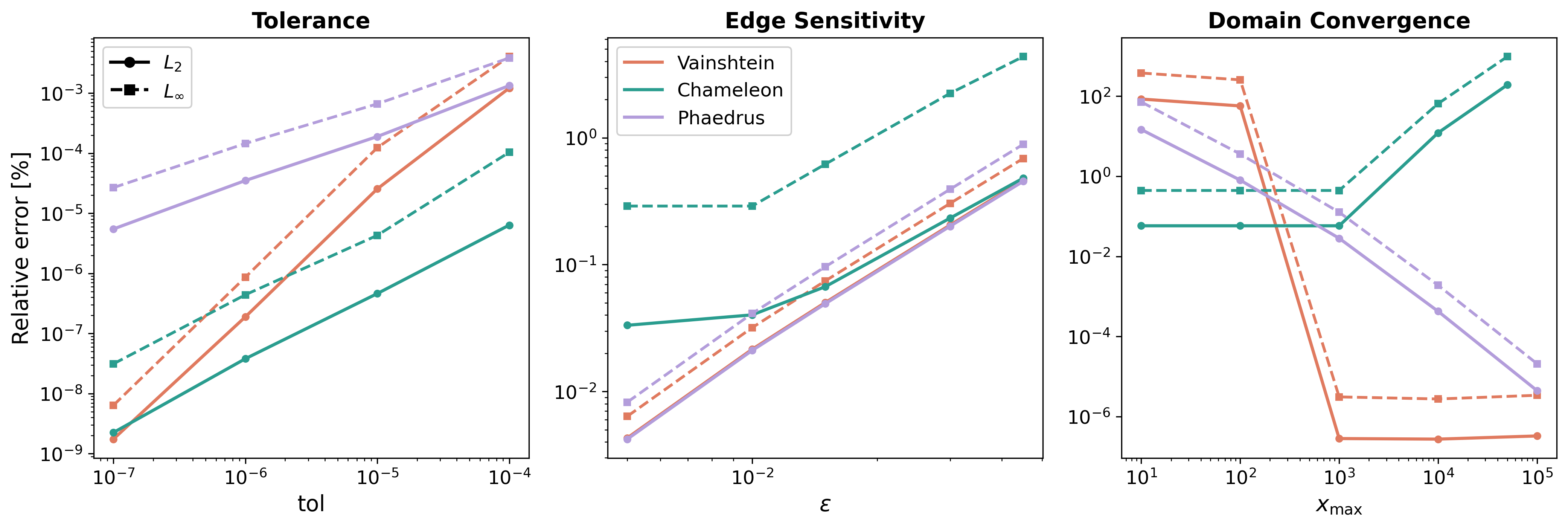}
    \caption{Numerical convergence and sensitivity tests for Vainshtein (\red{orange}), Chameleon (\red{teal}), and Phaedrus (\red{purple}). Circles and squares show relative $L_2$ and $L_\infty$ errors, respectively, with respect to a high-accuracy reference solution in each sweep. For derivative screenings (Vainshtein and Phaedrus) we compare the flux $x^2Q_x$, while for Chameleon we compare the field amplitude $Q$. \textbf{Left:} Convergence as a function of solver tolerance ($\mathrm{tol}$), showing the expected linear scaling for all mechanisms. \textbf{Middle:} Sensitivity to the edge-regularisation parameter ($\varepsilon$); smaller values correspond to a sharper source boundary. As expected, the Chameleon mechanism (relying on the thin-shell effect) is more sensitive to boundary sharpness. \textbf{Right:} Domain convergence as a function of the outer boundary $x_{\max}$. For derivative-based mechanisms, the interior solution becomes independent of the boundary placement for $x_{\max} \gtrsim 10^3$, while the Chameleon mechanism exhibits a characteristic rise in error at very large $x_{\max}$, arising from the numerical difficulty of resolving a localised exponential transition within a vast domain.}
    \label{fig:convergence}
\end{figure*}

To compute the numerical solutions shown in this work, we developed a publicly available dedicated solver for the screening equation, Eq.~\eqref{eq:scalar-pert-eom2}: \href{https://github.com/sergisl/escut}{\faGithub \;escut}. Here we summarise the numerical strategy, the explicit implementation equations, and the corresponding convergence tests. To facilitate exact reproducibility, the full Python notebooks used to generate the results in the main text and the tests presented in Fig.~\ref{fig:convergence} are included in the accompanying repository. In this work, the code is used to compute numerical solutions for the Vainshtein, Chameleon, and Phaedrus mechanisms. Furthermore, it provides a robust foundation for future extensions targeting mixed screening scenarios. For an alternative dedicated Vainshtein solver, see also \cite{Braden:2020zfa}.

In practice, we solve a dimensionless form of the master screening equation using the rescaled radial coordinate $x \equiv r/R$, where $R$ is the source radius:
\red{\begin{align}
&\frac{d}{dx}\left[x^2 \left(A + D Q\right) Q_x\right] + x^2 Q(B + CQ) \nonumber\\
&- \frac{d}{dx}\left[2F x Q_x^2\right] + (E - D) x^2 Q_x^2 + S x^2 \tilde{\rho}(x) = 0,
\end{align}
}
where $Q_x \equiv dQ/dx$, the parameters $\{A, B, C, D, E, F\}$ correspond directly to the theoretical coefficients from Eq.~\eqref{eq:scalar-pert-eom}, and $S \tilde{\rho}(x)$ represents the regularised density source profile. 

To pass this to the boundary-value solver, the equation is rewritten as a first-order system for the variables $(Q, Q_x)$. By expanding the derivatives and isolating the highest-order term, the principal equation dictates the second derivative:
\begin{align}
    Q_{xx} = \frac{\mathcal{N}}{\mathcal{D}},
\end{align}
with the numerator ($\mathcal{N}$) and denominator ($\mathcal{D}$) given by:
\begin{align}
    \mathcal{N} &= -S x^2 \tilde{\rho}(x) - 2x(A + D Q)Q_x - x^2 Q(B + C Q) \nonumber\\
    &\qquad+ 2 F Q_x^2 - E x^2 Q_x^2, \\
    \mathcal{D} &= x^2(A + D Q) - 4 F x Q_x.
\end{align}

This system is solved as a boundary-value problem on the domain $x\in[x_{\min},x_{\max}]$. The boundary conditions impose regularity at the origin, $Q_x(x_{\min})=0$, and a mixed (Robin) asymptotic condition at the outer boundary that matches the expected far-field decay toward $Q_\infty$. The corresponding decay scale is estimated from the linearised outer effective mass, while $Q_\infty$ sets the target field value at infinity (typically $Q_\infty=0$).

Given the model coefficients, source-profile parameters, and solver settings $(N,\mathrm{tol},x_{\min},x_{\max},\varepsilon)$, the code constructs an initial mesh and iteratively solves the nonlinear system using adaptive collocation until the requested tolerance is met. The parameter $N$ represents the initial number of mesh points in the log grid (which is adaptively refined up to a specified $\mathrm{max\_nodes}$). The tolerance ($\mathrm{tol}$) acts as a residual-based stopping criterion, therefore setting how close the numerical solution must satisfy the differential equation before the solver stops, with smaller $\mathrm{tol}$ resulting in stricter accuracies and longer runtimes. The parameter $\varepsilon$, defined in Eq.~\eqref{eq:eps-def}, represents the smoothing width for the source boundary at $x\sim1$; smaller values correspond to sharper, more physically realistic edges that are computationally harder to resolve. Finally, the code supports a $\mathrm{homotopy}$ continuation method, allowing the solver to approach the full highly nonlinear equation through a sequence of intermediate, well-behaved problems to guarantee convergence.

To assess the numerical reliability of this solver, we perform a set of convergence and sensitivity studies. Because closed-form solutions are generally unavailable in the nonlinear screened regime, convergence is evaluated by comparing each test solution to a high-accuracy reference solution computed with the most stringent numerical settings within that specific parameter sweep. For a given test solution $\mathcal{F}$ and reference solution $\mathcal{F}_{\mathrm{ref}}$, the pointwise error evaluated on a common set of sample points is $e_i = \left| \mathcal{F}(x_i) - \mathcal{F}_{\mathrm{ref}}(x_i) \right|$. We summarise this using the $L_2$ and $L_\infty$ norms:
\begin{align}
    L_2 &= \sqrt{\frac{1}{N_{\rm samp}} \sum_{i} |e_i|^2}, & L_\infty &= \max_{i} (|e_i|).
\end{align}
$L_2$ measures the global discrepancy, while $L_\infty$ tracks the largest local deviation, guaranteeing the absence of localised numerical artefacts. Note that for derivative screenings (Vainshtein and Phaedrus) we employ the physically relevant radial flux $\mathcal{F} \equiv x^2Q_x$ to evaluate convergence, while for Chameleon screening we employ the amplitude $\mathcal{F} \equiv Q$.

The results, shown in Fig.~\ref{fig:convergence}, confirm the stability of the solver. The tolerance study (left) shows consistent error reduction as $\mathrm{tol}$ is tightened. The $\varepsilon$ test (middle) confirms that the Chameleon mechanism is highly sensitive to the source boundary regularisation, consistent with the physics of the thin-shell effect. Finally, the domain-size study (right) demonstrates that Vainshtein and Phaedrus achieve interior insensitivity to the outer boundary for $x_{\max}\gtrsim10^3$. Conversely, the Chameleon error degrades for extremely large domains, reflecting the numerical difficulty of resolving a sharp, localised transition over a vast spatial grid without increasing $\mathrm{max\_nodes}$ to impractical levels. 

Overall, these tests demonstrate that the numerical solutions are stable and converged to sub-percent (or better) precision within the relevant interior regions. While we have utilised representative coefficient values for these baseline results, a critical next step is integrating this solver directly into the \texttt{Hi-COLA} framework, where these coefficients will be dynamically populated from $N$-body cosmological backgrounds.

\section{Linear and nonlinear coefficients}\label{app:expressions}
In this Appendix, we collect the full analytical expressions for the linear and nonlinear coefficients appearing in the perturbative metric and scalar equations of motion, as defined in Eqs.~\eqref{eq:E1}-\eqref{eq:S2}. To streamline the notation, we first define a set of $\gamma$ variables, which capture specific combinations of the background Hubble flow and the standard $\alpha$ parameters that appear repeatedly throughout the second-order expansion:
\begin{align}
    \gamma_E &= \frac{2\dot H-{\cal \tilde{E}}-{\cal \tilde{P}}}{H^2} = \frac{2\dot H+\tilde{\rho}+\tilde{p}}{H^2}, \\
    \gamma_M &= \frac{\dot\alpha_M}{H} + \alpha_M(3+\alpha_M), \\
    \gamma_K &= \frac{\dot\alpha_K}{H} + \alpha_K(3+\alpha_M), \\
    \gamma_B &= \frac{\dot\alpha_B}{H} + \alpha_B(3+\alpha_M) + \frac{\dot H}{H^2}\alpha_B, \\
    \gamma_A &= 2\alpha_M + \alpha_K + 6\alpha_B, \\
    \gamma_F &= \alpha_K - 2X\alpha_{KX}, \\
    \gamma_D &= \alpha_B - 2X\alpha_{BX}, \\
    \gamma_C &= 3\alpha_M + \alpha_K + 3\alpha_B - X\alpha_{KX}, \\
    \gamma_X &= 6\alpha_M + 2\alpha_K + \frac{81}{4}\alpha_B - 2X\alpha_{KX}\nonumber\\
    &\qquad- 27X\alpha_{BX} + 3X^2\alpha_{BXX}.
\end{align}

As discussed in the main text, restricting the landscape to luminal Horndeski theories guarantees that a significant fraction of the equation coefficients ($A_i, B_i, C_i, D_i$) identically vanish. Table~\ref{tab:linear_combined} and Table~\ref{tab:second_order_all} illustrate this sparse structure by cataloguing only the surviving non-zero entries for the linear and second-order terms, respectively, representing $150$ of the initial $300$ potential terms. 

The explicit algebraic expressions for these surviving coefficients are listed below, grouped by their corresponding equation of motion. These expressions were derived using a dedicated Mathematica package \href{https://github.com/sergisl/xAlpha}{\faGithub \;xAlpha}, which includes a reproducible notebook. To present the coefficients as compactly as possible, we adopt a hybrid notation, substituting the $\gamma$ variables defined above only when they strictly reduce the number of terms compared to the raw $\alpha$-basis. For the uncompressed derivations, we direct the reader to the accompanying repository.
Note that the package produces 3 versions of each coefficient: one written in the original $K$, $G_3$ and $G_4$ Horndeski functions, one purely in terms of the $\alpha$ coefficients, and one using the $\gamma$ coefficients defined above. For details on the full conversion of the coefficients here to the $\alpha$ and $\gamma$ language, we refer to the repository \href{https://github.com/sergisl/xAlpha}{\faGithub \;xAlpha}. Finally, also note that the package accommodates any of the 3 definitions for scalar perturbations discussed in Appendix~\ref{app:conversions}.

\begin{table}[ht!]
\centering
\renewcommand{\arraystretch}{1.25} 
\setlength{\tabcolsep}{10pt}
\begin{tabular}{@{}c l c c c@{}}
\toprule
Eq. & Term & $\Phi$ & $\Psi$ & $Q$ \\
\midrule
\multirow{3}{*}{$\mathcal{E}^{(1)}$} 
    & $Y^a$           & $A^{\Phi}_1$ & -- & $A^{Q}_1$ \\
    & $\dot Y^a$      & -- & $A^{\Psi}_2$ & $A^{Q}_2$ \\
    & $\nabla^2 Y^a$  & -- & \color{black} $A^\Psi_3$ & \color{black} $A^{Q}_3$ \\
\midrule
\multirow{2}{*}{$\mathcal{A}^{(1)}_i$} 
    & $\partial_iY^a$      & $B^{\Phi}_1$ & -- & $B^{Q}_1$ \\
    & $\partial_i \dot Y^a$ & -- & $B^{\Psi}_2$ & $B^{Q}_2$ \\
\midrule
\multirow{4}{*}{$\mathcal{P}^{(1)}_{ij}$} 
    & $\delta_{ij}Y^a$      & $C^{\Phi}_1$ & -- & $C^{Q}_1$ \\
    & $\delta_{ij}\dot Y^a$ & $C^{\Phi}_2$ & $C^{\Psi}_2$ & $C^{Q}_2$ \\
    & $\delta_{ij}\ddot Y^a$& -- & $C^{\Psi}_3$ & $C^{Q}_3$ \\
    & $\mathcal{D}_{ij} Y^a$& \color{black} $C^{\Phi}_4$ & \color{black} $C^{\Psi}_4$ & \color{black} $C^{Q}_4$ \\
\midrule
\multirow{4}{*}{$\mathcal{S}^{(1)}$} 
    & $Y^a$           & $D^{\Phi}_1$ & -- & \color{black} $D^{Q}_1$ \\
    & $\dot Y^a$      & $D^{\Phi}_2$ & $D^{\Psi}_2$ & $D^{Q}_2$ \\
    & $\ddot Y^a$     & -- & $D^{\Psi}_3$ & $D^{Q}_3$ \\
    & $\nabla^2 Y^a$  & \color{black} $D^{\Phi}_4$ & \color{black} $D^{\Psi}_4$ & \color{black} $D^{Q}_4$ \\
\bottomrule
\end{tabular}
\caption{Summary of the non-zero linear coefficients appearing in the first-order metric and scalar equations of motion. The leftmost column denotes the relevant equation ($\mathcal{E}, \mathcal{A}, \mathcal{P}, \mathcal{S}$), while the second column lists the specific differential operators applied to the field variables $Y^a \in \{\Phi, \Psi, Q\}$. The remaining columns map these operators to their respective fields. Blank entries (-) indicate coefficients that identically vanish within luminal Horndeski.}
\label{tab:linear_combined}
\end{table}

We first state the non-zero linear coefficients for the metric and scalar equations.
\paragraph{Linear metric $(00)$ equation}
\begin{flalign}
    &A_1^\Phi = 6 - \alpha_K - 6\alpha_B, &\\
    &A_2^\Psi = 6 - 3\alpha_B, \quad A_3^\Psi = 2, &\\
    &A_1^Q = 3\gamma_E - \frac{\dot H}{H^2}(\alpha_K + 6\alpha_B), &\\
    &A_2^Q = \alpha_K + 3\alpha_B, \quad A_3^Q = \alpha_B. &
\end{flalign}

\paragraph{Linear metric $(0i)$ equation}
\begin{flalign}
    &B_1^\Phi = -2 + \alpha_B, \quad B_2^\Psi = 2, &\\
    &B_1^Q = -\gamma_E + \frac{\dot H}{H^2}\alpha_B, \quad B_2^Q = \alpha_B. &
\end{flalign}

\paragraph{Linear metric $(ij)$ equation}
\begin{flalign}
    &C_1^\Phi = -6 + \gamma_E + \gamma_B - 2\alpha_M - \frac{4\dot H}{H^2}, &\\ 
    &C_2^\Phi = -2 + \alpha_B, \quad C_4^\Phi = 1, &\\
    &C_2^\Psi = -6 - 2\alpha_M, \quad C_3^\Psi = 2, \quad C_4^\Psi = -1, &\\
    &C_1^Q = \frac{\dot H}{H^2}\left[-6 + \gamma_E + \gamma_B - 2\alpha_M - \frac{2\dot H}{H^2}\alpha_B\right] \nonumber &\\
    &\qquad- \frac{\ddot H}{H^3}(2 - \alpha_B) + \frac{\mathcal{\dot{\tilde{P}}}}{H^3}, &\\
    &C_2^Q = -\gamma_E - \gamma_B + \frac{2\dot H}{H^2}\alpha_B, &\\
    &C_3^Q = \alpha_B, \quad C_4^Q = \alpha_M. &
\end{flalign}

\paragraph{Linear scalar equation}
\begin{flalign}
    &D_1^\Phi = -3\gamma_E + 3\gamma_B + \gamma_K + \frac{2\dot H}{H^2}(\alpha_K + 3\alpha_B), &\\ 
    &D_2^\Phi = \alpha_K + 3\alpha_B, \quad D_4^\Phi = \alpha_B, &\\
    &D_2^\Psi = -3\gamma_E + 3\gamma_B, \quad D_3^\Psi = 3\alpha_B, \quad D_4^\Psi = 2\alpha_M, &\\
    &D_1^Q = \frac{\dot H}{H^2}(-3\gamma_E + 3\gamma_B + \gamma_K) + \frac{\ddot H}{H^3}(\alpha_K + 3\alpha_B), \label{eq:D1Q} &\\
    &D_2^Q = -\gamma_K, \quad D_3^Q = -\alpha_K, &\\ 
    &D_4^Q = -\gamma_E + \gamma_B + 2\alpha_M - 2\alpha_B. \label{eq:D23Q} &
\end{flalign}

In the following pages, we state the non-zero nonlinear coefficients for the metric and scalar equations.

\begin{table*}[ht!]
\centering
\setlength{\tabcolsep}{10pt}
\renewcommand{\arraystretch}{1.3}
\begin{tabular}{@{}c l ccccccccc@{}}
\toprule
Eq. & Term & $\Phi\Phi$ & $\Psi\Psi$ & $QQ$ & $\Phi\Psi$ & $\Phi Q$ & $\Psi Q$ & $\Psi\Phi$ & $Q\Phi$ & $Q\Psi$ \\
\midrule
\multirow{6}{*}{$\mathcal{E}^{(2)}$}
    & $Y^a Y^b$          & $A^{\Phi\Phi}_1$ & -- & $A^{QQ}_1$ & -- & $A^{\Phi Q}_1$ & -- & \multicolumn{3}{c}{\emph{(sym)}} \\
    & $\dot Y^a\,Y^b$    & -- & $A^{\Psi\Psi}_2$ & $A^{QQ}_2$ & -- & -- & $A^{\Psi Q}_2$ & $A^{\Psi\Phi}_2$ & $A^{Q\Phi}_2$ & -- \\
    & $\dot Y^a\dot Y^b$ & -- & $A^{\Psi\Psi}_3$ & $A^{QQ}_3$ & -- & -- & $A^{\Psi Q}_3$ & \multicolumn{3}{c}{\emph{(sym)}} \\
    & $Y^a\nabla^2 Y^b$  & -- & $A^{\Psi\Psi}_4$ & $A^{QQ}_4$ & $A^{\Phi\Psi}_4$ & $A^{\Phi Q}_4$ & $A^{\Psi Q}_4$ & -- & -- & $A^{Q\Psi}_4$ \\
    & $\partial Y \partial Y$ & -- & $A^{\Psi\Psi}_5$ & $A^{QQ}_5$ & -- & -- & $A^{\Psi Q}_5$ & \multicolumn{3}{c}{\emph{(sym)}} \\
    & $\dot Y^a\nabla^2 Y^b$ & -- & -- & $A^{QQ}_6$ & -- & -- & -- & -- & -- & -- \\
\midrule
\multirow{5}{*}{$\mathcal{A}^{(2)}_i$}
    & $Y^a\partial_i Y^b$      & $B^{\Phi\Phi}_1$ & -- & $B^{QQ}_1$ & -- & $B^{\Phi Q}_1$ & -- & -- & $B^{Q\Phi}_1$ & -- \\
    & $\dot Y^a\partial_i Y^b$ & -- & $B^{\Psi\Psi}_2$ & $B^{QQ}_2$ & -- & -- & $B^{\Psi Q}_2$ & $B^{\Psi\Phi}_2$ & $B^{Q\Phi}_2$ & -- \\
    & $Y^a\partial_i\dot Y^b$  & -- & $B^{\Psi\Psi}_3$ & $B^{QQ}_3$ & -- & $B^{\Phi Q}_3$ & -- & -- & -- & $B^{Q\Psi}_3$ \\
    & $\dot Y^a\partial_i\dot Y^b$ & -- & -- & $B^{QQ}_4$ & -- & -- & -- & -- & -- & -- \\
    & $\partial Y \mathcal{D} Y$ & -- & -- & $B^{QQ}_5$ & -- & -- & -- & -- & -- & -- \\
\midrule
\multirow{10}{*}{$\mathcal{P}^{(2)}_{ij}$}
    & $\delta_{ij}Y Y$        & $C^{\Phi\Phi}_1$ & -- & $C^{QQ}_1$ & $C^{\Phi\Psi}_1$ & $C^{\Phi Q}_1$ & $C^{\Psi Q}_1$ & \multicolumn{3}{c}{\emph{(sym)}} \\
    & $\delta_{ij}Y \dot Y$   & $C^{\Phi\Phi}_2$ & -- & $C^{QQ}_2$ & $C^{\Phi\Psi}_2$ & $C^{\Phi Q}_2$ & $C^{\Psi Q}_2$ & $C^{\Psi\Phi}_2$ & $C^{Q\Phi}_2$ & $C^{Q\Psi}_2$ \\
    & $\delta_{ij}\dot Y \dot Y$& -- & $C^{\Psi\Psi}_3$ & $C^{QQ}_3$ & $C^{\Phi\Psi}_3$ & $C^{\Phi Q}_3$ & $C^{\Psi Q}_3$ & \multicolumn{3}{c}{\emph{(sym)}} \\
    & $\delta_{ij}Y \ddot Y$  & -- & -- & $C^{QQ}_4$ & $C^{\Phi\Psi}_4$ & $C^{\Phi Q}_4$ & $C^{\Psi Q}_4$ & -- & -- & $C^{Q\Psi}_4$ \\
    & $\delta_{ij}\dot Y \ddot Y$& -- & -- & $C^{QQ}_5$ & -- & -- & -- & -- & -- & -- \\
    & $\delta_{ij}\partial Y \partial Y$ & $C^{\Phi\Phi}_6$ & $C^{\Psi\Psi}_6$ & $C^{QQ}_6$ & -- & $C^{\Phi Q}_6$ & -- & \multicolumn{3}{c}{\emph{(sym)}} \\
    & $\delta_{ij}\partial Y \partial \dot Y$ & -- & -- & $C^{QQ}_7$ & -- & -- & -- & -- & -- & -- \\
    & $\partial_i Y \partial_j Y$ & $C^{\Phi\Phi}_8$ & $C^{\Psi\Psi}_8$ & $C^{QQ}_8$ & $C^{\Phi\Psi}_8$ & -- & -- & \multicolumn{3}{c}{\emph{(sym)}} \\
    & $\partial_i Y \partial_j \dot Y$ & -- & -- & $C_9^{QQ}$ & -- & -- & -- & -- & -- & -- \\
    & $Y \mathcal{D}_{ij}Y$   & $C^{\Phi\Phi}_{10}$ & $C^{\Psi\Psi}_{10}$ & $C^{QQ}_{10}$ & -- & -- & -- & -- & $C^{Q\Phi}_{10}$ & $C^{Q\Psi}_{10}$ \\
\midrule
\multirow{4}{*}{$\mathcal{S}^{(2)}$}
    & $Y^aY^b$        & $D^{\Phi\Phi}_1$ & -- & \color{black} $D^{QQ}_1$ & -- & $D^{\Phi Q}_1$ & -- & \multicolumn{3}{c}{\emph{(sym)}} \\
    & $Y^a\dot{Y}^b$        & $D^{\Phi\Phi}_2$ & $D^{\Psi\Psi}_2$ & $D^{QQ}_2$ & $D^{\Phi\Psi}_2$ & $D^{\Phi Q}_2$ & -- & -- & $D^{Q\Phi}_2$ & $D^{Q\Psi}_2$ \\
    & $\dot{Y}^a\dot{Y}^b$        & -- & $D^{\Psi\Psi}_3$ & $D^{QQ}_3$ & $D^{\Phi\Psi}_3$ & $D^{\Phi Q}_3$ & $D^{\Psi Q}_3$ & \multicolumn{3}{c}{\emph{(sym)}} \\
    & $Y^a\ddot{Y}^b$        & -- & $D^{\Psi\Psi}_4$ & $D^{QQ}_4$ & $D^{\Phi\Psi}_4$ & $D^{\Phi Q}_4$ & -- & -- & -- & $D^{Q\Psi}_4$ \\
    & $\dot{Y}^a\ddot{Y}^b$        & -- & -- & $D^{QQ}_5$ & -- & -- & $D^{\Psi Q}_5$ & -- & -- & $D^{Q\Psi}_5$ \\
    & $Y^a\nabla^2 Y^b$ & $D^{\Phi\Phi}_6$ & $D^{\Psi\Psi}_6$ & \color{black} $D^{QQ}_6$ & -- & $D^{\Phi Q}_6$ & $D^{\Psi Q}_6$ & $D^{\Psi\Phi}_6$ & $D^{Q\Phi}_6$ & $D^{Q\Psi}_6$ \\
    & $\partial Y \partial Y$ & $D^{\Phi\Phi}_7$ & $D^{\Psi\Psi}_7$ & \color{black} $D^{QQ}_7$ & $D^{\Phi\Psi}_7$ & -- & -- & \multicolumn{3}{c}{\emph{(sym)}} \\
    & $\dot{Y}^a\nabla^2 Y^b$        & -- & -- & $D^{QQ}_8$ & -- & $D^{\Phi Q}_8$ & $D^{\Psi Q}_8$ & -- & $D^{Q\Phi}_8$ & -- \\
    & $\partial \dot{Y} \partial Y$        & -- & -- & $D^{QQ}_9$ & -- & -- & $D^{\Psi Q}_9$ & -- & $D^{Q\Phi}_9$ & -- \\
    & $\ddot{Y}^a\nabla^2 Y^b$        & -- & -- & $D^{QQ}_{10}$ & -- & -- & -- & -- & -- & -- \\
    & $\partial \dot{Y} \partial \dot{Y}$ & -- & -- & $D^{QQ}_{11}$ & -- & -- & -- & \multicolumn{3}{c}{\emph{(sym)}} \\
    & $\mathcal{D}_{ij}Y\partial^i\partial^jY$ & -- & -- & \color{black} $D^{QQ}_{12}$ & -- & -- & -- & \multicolumn{3}{c}{\emph{(sym)}} \\
\bottomrule
\end{tabular}
\caption{Summary of the non-zero quadratic coefficients appearing in the second-order metric and scalar equations of motion. Rows denote the specific nonlinear derivative operators applied to the interacting field pairs $Y^a Y^b$. The columns map to the respective field combinations, with symmetric cross-terms (e.g. $\Phi\Psi$ and $\Psi\Phi$) only counted once. As with the linear case, blank entries (-) indicate interactions that are strictly zero in luminal Horndeski.}
\label{tab:second_order_all}
\end{table*}

\begin{widetext}
\paragraph{Quadratic metric $(00)$ equation}
\begin{flalign}
    &A_1^{\Phi\Phi} = -\gamma_X, &\\
    &A_4^{\Phi\Psi} = 2, \quad A_2^{\Psi\Phi} = -3\gamma_D, &\\
    &A_2^{\Psi\Psi} = 12 - 6\alpha_B, \quad A_3^{\Psi\Psi} = 1, \quad A_4^{\Psi\Psi} = 4, \quad A_5^{\Psi\Psi} = 6, &\\
    &A_1^{\Phi Q} = 2( 6\gamma_E - 6\gamma_B - \gamma_K + 3\gamma_A) - \frac{2\dot H}{H^2}(2\gamma_A + \gamma_X - 4\alpha_M - 6\alpha_B), &\\
    &A_4^{\Phi Q} = \frac{\gamma_D}{2} + \alpha_B, \quad A_2^{Q\Phi} = -3\gamma_D + \gamma_X, &\\
    &A_2^{\Psi Q} = -3( \gamma_B - 2\alpha_M - 3\alpha_B) - \frac{3\dot H}{H^2}(\gamma_D - \alpha_B), \quad
    A_3^{\Psi Q} = 2X\alpha_{BX}, &\\
    &A_4^{\Psi Q} = \alpha_B, \quad A_5^{\Psi Q} = -2\alpha_B, \quad A_4^{Q\Psi} = \alpha_M, &\\
    &A_1^{QQ} = -3\Biggl\lceil 3\gamma_E + \frac{\mathcal{\dot{\tilde{P}}}}{H^3} - \frac{\dot H}{H^2}\left[6 + \gamma_E - \frac{2}{3}\gamma_K - 4\gamma_B + 2\gamma_A - \frac{\dot H}{H^2}\left(\frac{1}{3}\gamma_X + \alpha_K\right)\right] \nonumber\\
    &\qquad - \frac{\ddot H}{H^3}(2 - \alpha_B) + \frac{\ddot\phi}{H \dot\phi}\left(\gamma_E - \frac{\dot H}{H^2}\alpha_B\right) - \frac{1}{3}\left[\left(\frac{\ddot \phi}{H\dot\phi}\right)^2 - \frac{\dddot\phi}{H^2\dot\phi} + 2\frac{\dot H\ddot\phi}{H^3\dot\phi}\right](\alpha_K + 3\alpha_B)\Biggr\rfloor,&\\
    &A_2^{QQ} = 3\gamma_B + \gamma_K - 3\alpha_K - 9\alpha_B - \frac{\dot H}{H^2}(3\gamma_D - \gamma_X - \alpha_K) - \frac{\ddot\phi}{H\dot\phi}(\alpha_K + 3\alpha_B), \quad
    A_3^{QQ} = -\gamma_D + \frac{1}{6}\gamma_X - \frac{1}{6}\alpha_K, &\\
    &A_4^{QQ} = \frac{1}{2}\gamma_B - \frac{3}{2}\alpha_B + \frac{\dot H}{2H^2}\gamma_D - \frac{\ddot\phi}{2H\dot\phi}\alpha_B, \quad
    A_5^{QQ} = 2\gamma_E + \alpha_K + 2\alpha_B - \frac{2\ddot\phi}{H\dot\phi}\alpha_B, \quad A_6^{QQ} = \gamma_D. &
\end{flalign}

\paragraph{Quadratic metric $(0i)$ equation}
\begin{flalign}
    &B_1^{\Phi\Phi} = 4 + \gamma_D - 2\alpha_B, &\\
    &B_2^{\Psi\Phi} = -2, \quad B_2^{\Psi\Psi} = 4, \quad B_3^{\Psi\Psi} = 4, &\\
    &B_1^{\Phi Q} = -\gamma_E + \gamma_B - \gamma_A + \frac{\dot H}{H^2}\gamma_D, \quad B_3^{\Phi Q} = \gamma_D, &\\
    &B_1^{Q\Phi} = \gamma_B - 2\alpha_M - 3\alpha_B - \frac{2\dot H}{H^2}X\alpha_{BX}, \quad B_2^{Q\Phi} = -2X\alpha_{BX}, &\\
    &B_2^{\Psi Q} = 2\alpha_M + 3\alpha_B, \quad B_3^{Q\Psi} = 2\alpha_M, &\\
    &B_1^{QQ} = 3\gamma_E - \frac{\dot H}{H^2}(6 - 2\gamma_B + 4\alpha_M + \alpha_K + 9\alpha_B) + \frac{\dot H^2}{H^4}\gamma_D - \frac{2\ddot H}{H^3} + \frac{\mathcal{\dot{\tilde{P}}}}{H^3} + \frac{\ddot\phi}{H\dot\phi}\gamma_E &\\
    &\qquad - \left[\left(\frac{\ddot \phi}{H\dot\phi}\right)^2 - \frac{\dddot\phi}{H^2\dot\phi} + 2\frac{\dot H\ddot\phi}{H^3\dot\phi}\right]\alpha_B, \quad
    B_2^{QQ} = \gamma_B - \alpha_K - 3\alpha_B + \frac{\dot H}{H^2}\gamma_D - \frac{\ddot\phi}{H\dot\phi}\alpha_B, &\\
    &B_3^{QQ} = \gamma_B - 3\alpha_B + \frac{\dot H}{H^2}\gamma_D - \frac{\ddot\phi}{H\dot\phi}\alpha_B, \quad
    B_4^{QQ} = \gamma_D, \quad B_5^{QQ} = \alpha_M + \alpha_B. &
\end{flalign}

\paragraph{Quadratic metric $(ij)$ equation}
\begin{flalign}
    &C_2^{\Phi\Phi} = 8 + \gamma_D - 4\alpha_B, \quad C_6^{\Phi\Phi} = 1, \quad C_8^{\Phi\Phi} = -1, \quad C_{10}^{\Phi\Phi} = 2, &\\
    &C_1^{\Phi\Psi} = 2\left(-6 + \gamma_E + \gamma_B - 2\alpha_M - \frac{4\dot H}{H^2}\right), \quad
    C_2^{\Phi\Psi} = 4(3 + \alpha_M), \quad C_3^{\Phi\Psi} = -2, \quad C_4^{\Phi\Psi} = 4, \quad C_8^{\Phi\Psi} = 2, &\\
    &C_2^{\Psi\Phi} = 2(2 - \alpha_B), \quad C_3^{\Psi\Psi} = 1, \quad C_6^{\Psi\Psi} = 2, \quad C_8^{\Psi\Psi} = -3, \quad C_{10}^{\Psi\Psi} = 2, &\\
    &C_1^{\Phi\Phi} = \frac{1}{2}\left[-24 + 3\gamma_E + 5\gamma_B - \gamma_A - 3\gamma_D - 8\alpha_M - \gamma_D\alpha_M - \frac{\dot\gamma_D}{H} - \frac{\dot H}{H^2}(16 + \gamma_D)\right], &\\
    &C_1^{\Phi Q} = 3\gamma_E + 2\gamma_M - \gamma_B\alpha_M - \frac{\dot\gamma_B}{H} + \frac{\mathcal{\dot{\tilde{P}}}}{H^3} - \frac{\dot H}{H^2}\left[6 - \gamma_E - \gamma_B + 3\gamma_D + \gamma_D\alpha_M + \alpha_K + 6\alpha_B\right] \nonumber\\
    &\qquad - \frac{\dot H\dot\gamma_D}{H^3} + \frac{\dot H^2}{H^4}(\gamma_D - 4\alpha_B) - \frac{\ddot H}{H^3}(2 + \gamma_D - 3\alpha_B) + \frac{\dddot\phi}{H^2\dot\phi}(\gamma_D - \alpha_B),&\\
    &C_2^{\Phi Q} = \gamma_E + 3\gamma_B - \gamma_A -\gamma_D ( 3 + \alpha_M) - \frac{\dot\gamma_D}{H} + \frac{\dot H}{H^2}(\gamma_D - 4\alpha_B), \quad
    C_3^{\Phi Q} = -2X\alpha_{BX}, \quad C_4^{\Phi Q} = \alpha_B + 2X\alpha_{BX}, &\\
    &C_6^{\Phi Q} = -\alpha_M - \frac{1}{2}\alpha_B, \quad C_8^{\Phi Q} = 2(\alpha_M + \alpha_B), \quad
    C_2^{Q\Phi} = \gamma_B - 2\alpha_M - 3\alpha_B + \frac{\dot H}{H^2}(\gamma_D - \alpha_B), \quad C_{10}^{Q\Phi} = -\alpha_M, &\\
    &C_1^{\Psi Q} = 2 \left[ \frac{\dot H}{H^2}( - 6 + \gamma_E + \gamma_B - 2\alpha_M) - \frac{\ddot H}{H^3}(2-\alpha_B) + \frac{\mathcal{\dot{\tilde{P}}}}{H^3} - \frac{2\dot H^2}{H^4}\alpha_B \right], \quad
    C_2^{\Psi Q} = 2\left(\gamma_E + \gamma_B - \frac{\dot H}{H^2}\alpha_B\right), &\\
    &C_3^{\Psi Q} = 2\alpha_M, \quad C_4^{\Psi Q} = 2\alpha_B, \quad C_8^{\Psi Q} = 2\alpha_M, \quad
    C_2^{Q\Psi} = -2\gamma_M, \quad C_4^{Q\Psi} = -2\alpha_M, \quad C_{10}^{Q\Psi} = \alpha_M, &\\
    &C_1^{QQ} = \frac{\dot H}{H^2}(3\gamma_E + 2\gamma_M - \gamma_B\alpha_M) - \frac{\dot H\dot\gamma_B}{H^3} - \frac{\mathcal{\ddot{\tilde{P}}}}{2H^4} + \frac{\mathcal{\dot{\tilde{P}}}}{H^3}\left(\frac{\dot H}{H^2} + \frac{\ddot\phi}{2H\dot\phi}\right) - \frac{\dot H^2\dot\gamma_D}{2H^5} + \frac{3\dot H^3}{2H^6}\gamma_D \nonumber\\
    &\quad - \frac{\dot H^2}{H^4}\left(3 - \gamma_B + \frac{1}{2}\gamma_A + \frac{3}{2}\gamma_D + \alpha_M  + 6\alpha_B + \frac{1}{2}\gamma_D\alpha_M\right) + \frac{\ddot H}{H^3}(3 - \gamma_B + 3\alpha_M + 3\alpha_B) &\\
    &\quad- \frac{\dot H\ddot H}{H^5}(2 + \gamma_D - \alpha_B) - \frac{\dot H\ddot\phi}{H^3\dot\phi}(3 - \gamma_E - \gamma_B + \alpha_M) - \frac{\ddot H\ddot\phi}{H^4\dot\phi}(1 - \alpha_B) + \left(\frac{\ddot\phi^2}{4H^2X} - \frac{\dddot\phi}{2H^2\dot\phi}\right)(\gamma_E + \gamma_B) \nonumber\\
    &\quad - \frac{\dot H\dddot\phi}{H^4\dot\phi}(\gamma_D + \alpha_B) - \left(\frac{3\dot H^2\ddot\phi}{H^5\dot\phi} + \frac{\dot H\ddot\phi^2}{H^4X} + \frac{\ddot\phi^3}{2H^3X\dot\phi} - \frac{3\ddot\phi\dddot\phi}{4H^3X}\right)\alpha_B,&\\
    &C_2^{QQ} = 3\gamma_E - \gamma_B\alpha_M - \frac{\dot\gamma_B}{H} + \frac{\mathcal{\dot{\tilde{P}}}}{H^3} - \frac{\dot H}{H^2}\left(6 - 2\gamma_B + 2\gamma_A + 3\gamma_D + \gamma_D\alpha_M - \alpha_K\right) - \frac{2\ddot H}{H^3} - \frac{\dot H\dot\gamma_D}{H^3}\nonumber\\
    &\quad + \frac{\ddot\phi}{H\dot\phi}(\gamma_E + \gamma_B) + \left(\frac{3\dot H^2}{H^4} - \frac{\ddot H}{H^3} + \frac{\dddot\phi}{H^2\dot\phi}\right)\gamma_D + \left(\frac{\ddot H}{H^3} - \frac{4\dot H\ddot\phi}{H^3\dot\phi} - \frac{\ddot\phi^2}{H^2X} + \frac{\dddot\phi}{H^2\dot\phi}\right)\alpha_B,&\\
    &C_3^{QQ} = -\frac{1}{2}\gamma_A + \gamma_B - \frac{3}{2}\gamma_D + \alpha_M - \frac{1}{2}\gamma_D\alpha_M - \frac{\dot\gamma_D}{2H} + \frac{3\dot H}{2H^2}\gamma_D + \frac{\ddot\phi}{H\dot\phi}\alpha_B, &\\
    &C_4^{QQ} = -\gamma_B + 3\alpha_B - \frac{\dot H}{H^2}\gamma_D + \frac{\ddot\phi}{H\dot\phi}\alpha_B, \quad C_5^{QQ} = \gamma_D, \quad C_7^{QQ} = \alpha_M + \alpha_B, &\\
    &C_6^{QQ} = \frac{1}{2}\gamma_E + \frac{1}{2}\gamma_B - \alpha_B - \frac{2\dot H}{H^2}(\alpha_M + \alpha_B) + \frac{\ddot\phi}{H\dot\phi}\alpha_M, &\\
    &C_8^{QQ} = -\gamma_E + \alpha_M + \frac{2\dot H}{H^2}(\alpha_M + \alpha_B) - \frac{\ddot\phi}{\dot\phi}\alpha_M, \quad C_9^{QQ} = \alpha_M + \alpha_B, \quad C_{10}^{QQ} = -\alpha_M^2 - \frac{\dot\phi}{H}\alpha_{M\phi}. &
\end{flalign}

\paragraph{Quadratic scalar equation}
\begin{flalign}
    &D_1^{\Phi\Phi} = \frac{1}{2}\Biggl\lceil 9(\gamma_E - \gamma_B + 2\gamma_D) - 4\gamma_K + 3\gamma_F + - 6\alpha_M + \gamma_X\alpha_M - 3\gamma_D\alpha_M - 18\alpha_B \nonumber\\
    &\quad + \frac{\dot\gamma_X-3\dot\gamma_D}{H} + \frac{\dot H}{H^2}\left(\gamma_X + \gamma_F + 6\gamma_D - 7\alpha_K - 24\alpha_B\right) + \frac{7\ddot\phi}{H\dot\phi}(-\gamma_X + \gamma_F + 9\gamma_D + \alpha_K)\Biggr\rfloor, &\\
    &D_2^{\Phi\Phi} = \frac{1}{3}\gamma_F + 2\gamma_D - \alpha_K - 4\alpha_B, \quad D_6^{\Phi\Phi} = -\alpha_B - 2X\alpha_{BX}, \quad D_7^{\Phi\Phi} = -2X\alpha_{BX}, &\\
    &D_2^{\Phi\Psi} = \gamma_E - \gamma_B - \gamma_A + \left(3 + \alpha_M + \frac{\dot H}{H^2}\right)\gamma_D + \frac{\dot\gamma_D}{H}, &\\
    &D_3^{\Phi\Psi} = -2X\alpha_{BX}, \quad D_4^{\Phi\Psi} = - \alpha_B -2X\alpha_{BX}, \quad D_7^{\Phi\Psi} = -\alpha_B, \quad D_6^{\Psi\Phi} = 2\alpha_B, &\\
    &D_2^{\Psi\Psi} = 2(\gamma_B-\gamma_E), \quad D_3^{\Psi\Psi} = -\alpha_M - \alpha_B, \quad D_4^{\Psi\Psi} = 2\alpha_B, \quad D_6^{\Psi\Psi} = 8\alpha_M, \quad D_7^{\Psi\Psi} = 3\alpha_M, &\\
    &D_1^{\Phi Q} = 9\gamma_E + \gamma_K\alpha_M + 3\gamma_B\alpha_M + \frac{\dot\gamma_K+3\dot\gamma_B}{H} + \frac{3\mathcal{\dot{\tilde{P}}}}{H^3} - \frac{6\ddot H}{H^3} + \frac{\dot H(\dot\gamma_F + 6\dot\gamma_D)}{H^3} + \frac{\dot H^2}{H^4}(3\gamma_D + 2\alpha_K)\nonumber\\
    &\quad + \frac{\dot H}{H^2}\left[ -18 + 12\gamma_B + 4\gamma_K + 3\gamma_F + 18\gamma_D - 6\gamma_A + \gamma_F\alpha_M + 6\gamma_D\alpha_M - 3\alpha_K \right] \nonumber\\
    &\quad - \frac{\ddot\phi}{H\dot\phi}( -3\gamma_E + 3\gamma_B + \gamma_K ) - \frac{2\dot H\ddot\phi}{H^3\dot\phi}(\alpha_K + 3\alpha_B) + \left(\frac{\dddot\phi}{H^2\dot\phi} - \frac{\ddot H}{H^3}\right)(-\gamma_F - 3\gamma_D + \alpha_K + 3\alpha_B),&\\
    &D_2^{\Phi Q} = \frac{1}{3}\gamma_K - \gamma_F - 3\gamma_D - \frac{1}{3}\gamma_F\alpha_M - \gamma_D\alpha_M - \frac{\dot\gamma_F}{3H} - \frac{\dot\gamma_D}{H}, \quad D_3^{\Phi Q} = -\gamma_D - \frac{1}{3}\gamma_F, &\\
    &D_4^{\Phi Q} = -\frac{1}{3}\gamma_F - \gamma_D + \frac{1}{3}\alpha_K, \quad D_5^{\Phi Q} = \gamma_D, \quad D_6^{\Phi Q} = -\gamma_E + \gamma_B - \gamma_A + \left(1+\alpha_M+ \frac{\dot H}{H^2}\right)\gamma_D + \frac{\dot\gamma_D}{H}, &\\
    &D_7^{\Phi Q} = -\gamma_E + \gamma_B - \frac{2}{3}\alpha_M - \frac{5}{6}\alpha_K - 6\alpha_B + \frac{\dot H}{H^2}\gamma_D, \quad D_8^{\Phi Q} = -\gamma_D, &\\
    &D_3^{\Psi Q} = -3\gamma_D - \gamma_D\alpha_M + \alpha_K - \left(\frac{\gamma_D}{H}\right)^{\boldsymbol{\cdot}}, \quad D_5^{\Psi Q} = \gamma_D, \quad D_6^{\Psi Q} = -2(\gamma_E - \gamma_B - 2\alpha_M + 2\alpha_B), &\\
    &D_8^{\Psi Q} = 4(\alpha_M + \alpha_B), \quad D_9^{\Psi Q} = 2(\alpha_M + \alpha_B), &\\
    &D_2^{Q\Phi} = \gamma_B + \frac{1}{3}\gamma_K - \left(1 + \frac{\ddot\phi}{H\dot\phi}\right)(\alpha_K + 3\alpha_B) + \frac{\dot H}{H^2}\left(\frac{1}{3}\gamma_F + 2\gamma_D + \frac{2}{3}\alpha_K\right), &\\
    &D_6^{Q\Phi} = \gamma_B - 3\alpha_B + \frac{\dot H}{H^2}\gamma_D - \frac{\ddot\phi}{H\dot\phi}\alpha_B, \quad D_8^{Q\Phi} = \gamma_D, \quad D_9^{Q\Phi} = \gamma_D, &\\
    &D_2^{Q\Psi} = 3\gamma_E + \gamma_B\alpha_M + \frac{\dot\gamma_B}{H} + \frac{\mathcal{\dot{\tilde{P}}}}{H^3} - \frac{\dot H}{H^2}\left(6 - 2\gamma_B + \gamma_A - 3\gamma_D + 2\alpha_M - \gamma_D\alpha_M\right) \nonumber\\
    &\quad + \frac{\dot H}{H^2}\left(\frac{\gamma_D}{H}\right)^{\boldsymbol{\cdot}} - \frac{\ddot H}{H^3}(2 - \gamma_D + \alpha_B) + \frac{\ddot\phi}{H\dot\phi}(\gamma_E - \gamma_B) - \frac{\dddot\phi}{H^2\dot\phi}(\gamma_D - \alpha_B),&\\
    &D_4^{Q\Psi} = \gamma_B - \left(3+\frac{\ddot\phi}{H\dot\phi}\right)\alpha_B + \frac{\dot H}{H^2}\gamma_D, \quad D_5^{Q\Psi} = \gamma_D, \quad D_6^{Q\Psi} = 2\left[\gamma_M - 3\alpha_M + \left(\frac{2\dot H}{H^2} - \frac{2\ddot\phi}{H\dot\phi}\right)\alpha_M\right], &\\
    &D_2^{QQ} = \frac{1}{3}\Bigg[ -\gamma_K\alpha_M - \frac{X}{H^4}\left(\frac{\gamma_K H^3}{X}\right)^{\boldsymbol{\cdot}} - \frac{\dot H}{H^2}(\gamma_F + 3\gamma_D)(3+\alpha_M) - \frac{\dot H}{H}\left(\frac{\gamma_F + 3\gamma_D}{H^2}\right)^{\boldsymbol{\cdot}} \nonumber\\
    &\quad + 2\left(\frac{\dddot\phi}{H^2\dot\phi} - \frac{\ddot H}{H^3}\right)(\alpha_K - X\alpha_{KX}) + \alpha_K\left(\frac{6\dot H^2}{H^4} - \frac{2\dot H\ddot\phi}{H^3\dot\phi} - \frac{\ddot\phi^2}{H^2X}\right) \Bigg],&\\
    &D_3^{QQ} = \frac{1}{6}\left[ \alpha_K\left(6 + \alpha_M - \frac{4\dot H}{H^2} + \frac{2\ddot\phi}{H\dot\phi}\right) - 2X\alpha_{KX}(3+\alpha_M) - \frac{3\dot H}{H^2}\gamma_D + H\left(\frac{\gamma_F}{H^2}\right)^{\boldsymbol{\cdot}} \right],&\\
    &D_4^{QQ} = \frac{1}{3}\left[ -\gamma_K + \alpha_K\left(3 - \frac{4\dot H}{H^2} + \frac{2\ddot\phi}{H\dot\phi}\right) - \frac{\dot H}{H^2}(3\gamma_D - 2X\alpha_{KX}) \right], \quad D_5^{QQ} = -\frac{2}{3}(\alpha_K-X\alpha_{KX}), &\\
    &D_6^{QQ} = 3\gamma_E + 2\gamma_M - \gamma_B(2 - \alpha_M) + 6\alpha_B - 6\alpha_M + \frac{\mathcal{\dot{\tilde{P}}}}{H^3} + \frac{X}{H^4} \left( \frac{\gamma_B H^3}{X} \right)^{\boldsymbol{\cdot}} \nonumber\\
    &\quad - \frac{\dot H}{H^2}\big[ 6 + \gamma_E + \alpha_K + 8\alpha_B - \gamma_D(1+\alpha_M) \big] + \frac{\dot H}{H^2}\left(\frac{\gamma_D}{H}\right)^{\boldsymbol{\cdot}} \nonumber\\
    &\quad + \frac{2\ddot\phi}{H\dot\phi}(\gamma_E - 2\alpha_M + 2\alpha_B) - \frac{2\ddot H}{H^3} + 2X\alpha_{BX}\left(\frac{\dddot\phi}{H^2\dot\phi} - \frac{\ddot H}{H^3}\right),\label{eq:D6QQ}&\\
    &D_7^{QQ} = \frac{1}{2}(\gamma_E + \gamma_K - \gamma_B) - \alpha_K + \alpha_B - \alpha_M - \frac{\dot H}{H^2}(2\gamma_E - 2\gamma_B + \gamma_A - 4\alpha_M) + \frac{\dot H^2}{H^4}\gamma_D \nonumber\label{eq:D7QQ}&\\
    &\quad + \frac{\ddot\phi}{H\dot\phi}(\gamma_E - \gamma_B - 2\alpha_M + 2\alpha_B), \quad D_8^{QQ} = -\gamma_E + \gamma_B + \gamma_D(1 + \alpha_M) + 2\alpha_M - \alpha_K - 2\alpha_B + \left(\frac{\gamma_D}{H}\right)^{\boldsymbol{\cdot}},&\\
    &D_9^{QQ} = -\gamma_E + \gamma_B + 2\alpha_M - \alpha_K - 2\alpha_B + \frac{\dot H}{H^2}\gamma_D, \quad D_{10}^{QQ} = \gamma_D, \quad D_{11}^{QQ} = \gamma_D, \quad D_{12}^{QQ} = 2(\alpha_M + \alpha_B). &
\end{flalign}
\end{widetext}

Note that in the main text, we define the Phaedrus operators as $\kappa_{-}\equiv D_6^{QQ}$ and $\kappa_{+}\equiv D_7^{QQ}$. We here isolate these two specific coefficients and evaluate them for the reduced Lagrangian $\mathcal{L}_\phi=G_4(\phi)R+K(\phi,X)$. Written explicitly in terms of $K$-derivatives, they take the form:
\begin{align}
    \kappa_{-} &= -\frac{2X}{H^3M_*^2}\left[ H \left(K_{\phi X} + K_{XX}\ddot{\phi}\right) - \dot{H}\dot{\phi}K_{XX} \right],\nonumber\\
    \kappa_{+} &= -\frac{X}{H^3M_*^2}\Big[ (H^2 - 4\dot{H})\dot{\phi}K_{XX} \nonumber\\
    &+ H \Big( K_{\phi X} + 2X K_{\phi XX} + \ddot{\phi}(5K_{XX} + 2X K_{XXX}) \Big) \Big].
    \label{eq:d2d3-reduced-K}
\end{align}

Finally, we note that the coefficient $D_1^{QQ}$ was omitted from the scalar equation lists above, as its length and complexity require special treatment. This coefficient dictates the nonlinear contribution to the effective scalar mass ($M_{\rm nl}^2$), making it the fundamental mathematical driver of the Chameleon screening mechanism. We decompose this coefficient into a purely shift-symmetric contribution ($D_{1SS}$) and a non-shift-symmetric remainder ($\mathcal{M}_\phi$):
\begin{align}
    D_1^{QQ}=D_{1SS} + \frac{1}{H^2}\mathcal{M}_{\phi}^2,\label{eq:D1QQ}
\end{align}

It is crucial to highlight the physical origin of the pure shift-symmetric term. As discussed in the main text, for strictly shift-symmetric theories, a bare mass term for the standard perturbation $\delta\phi$ is explicitly forbidden. However, by transforming to the dimensionless variable $Q \equiv H\delta\phi/\dot\phi$, a shift-symmetric effective mass is generated. This contribution arises entirely from the background time evolution, rather than a potential. To express this shift-symmetric contribution compactly, we first define the following effective combinations of the Horndeski functions:

\begin{align}
\mathcal{K}_1 &\equiv \frac{X^2}{M_*^2 H^2} (3 K_{XX} + 2 X K_{XXX}), \\ \mathcal{K}_2 &\equiv \frac{X^2}{M_*^2 H^2} (3 K_{XX} + 12 X K_{XXX} + 4 X^2 K_{XXXX}) \\
\mathcal{B}_1 &\equiv \frac{X \dot{\phi}}{M_*^2 H} (G_{3X} + 5 X G_{3XX} + 2 X^2 G_{3XXX}), \\ \mathcal{B}_2 &\equiv \frac{X \dot{\phi}}{M_*^2 H} (12 X G_{3XX} + 18 X^2 G_{3XXX} + 4 X^3 G_{3XXXX})
\end{align}
Using these auxiliary functions, the purely shift-symmetric contribution evaluates to:
\begin{widetext}
\begin{align}
D_{1SS} = \,\, & 2 \left( \frac{\dot{H}}{H^2} - \frac{\ddot{\phi}}{H\dot{\phi}} \right) \Bigg\{ \left[ \mathcal{K}_1 \frac{\dot{H}}{H^2} - (\mathcal{K}_2 + 3\mathcal{B}_2) \frac{\ddot{\phi}}{H\dot{\phi}} \right] \left( \frac{\dot{H}}{H^2} - \frac{\ddot{\phi}}{H\dot{\phi}} \right) \nonumber \\
& \qquad\qquad\qquad\qquad - (\mathcal{K}_1 + 3\mathcal{B}_1) \left[ 3 \left( 1 - \frac{\dot{H}}{H^2} \right) \left( \frac{\dot{H}}{H^2} - \frac{\ddot{\phi}}{H\dot{\phi}} \right) + 2 \left( \frac{\ddot{H}}{H^3} - \frac{\dddot{\phi}}{H^2\dot{\phi}} \right) \right] \Bigg\}.
\end{align}
\end{widetext}
The remaining non-shift-symmetric contribution, $\mathcal{M}_\phi$, contains all explicit scalar field derivatives of the Horndeski functions (e.g. $K_{\phi\phi\phi}$, $G_{3\phi X}$). Because these terms explicitly break shift symmetry, $\mathcal{M}_\phi$ identically vanishes in the exact shift-symmetric limit. It takes the form:
\begin{widetext}
\begin{align}
\mathcal{M}_\phi^2 = -\frac{\dot{\phi} X}{2 H M_\star^2} \Bigg[
& -12 G_{4\phi\phi\phi} \left(2 + \frac{\dot{H}}{H^2}\right) - \frac{2}{H^2} \Big[ K_{\phi\phi\phi} - 2X (K_{\phi\phi\phi X} - G_{3\phi\phi\phi\phi}) \Big] \nonumber \\
& + 2 (K_{\phi X} - 2G_{3\phi\phi}) \left( -6 \frac{\dot{H}}{H^2} + 5 \frac{\dot{H}^2}{H^4} - 2 \frac{\ddot{H}}{H^3} + 6 \frac{\ddot{\phi}}{H\dot{\phi}} - 6 \frac{\dot{H}\ddot{\phi}}{H^3\dot{\phi}} + \frac{\ddot{\phi}^2}{H^2\dot{\phi}^2} + 2 \frac{\dddot{\phi}}{H^2\dot{\phi}} \right) \nonumber \\
& + 8 X (K_{\phi XX} - G_{3\phi\phi X}) \left( -3 \frac{\dot{H}}{H^2} + 4 \frac{\dot{H}^2}{H^4} - \frac{\ddot{H}}{H^3} + 3 \frac{\ddot{\phi}}{H\dot{\phi}} - 9 \frac{\dot{H}\ddot{\phi}}{H^3\dot{\phi}} + 5 \frac{\ddot{\phi}^2}{H^2\dot{\phi}^2} + \frac{\dddot{\phi}}{H^2\dot{\phi}} \right) \nonumber \\
& + 8 X^2 (K_{\phi XXX} - G_{3\phi\phi XX}) \left( \frac{\dot{H}^2}{H^4} - 4 \frac{\dot{H}\ddot{\phi}}{H^3\dot{\phi}} + 3 \frac{\ddot{\phi}^2}{H^2\dot{\phi}^2} \right) + 24 X^2 G_{3\phi\phi XX} \left( -2 \frac{\dot{H}}{H^2} + 3 \frac{\ddot{\phi}}{H\dot{\phi}} \right) \nonumber \\
& + 2 (K_{\phi\phi X} - 2G_{3\phi\phi\phi}) \left( 3 \frac{\dot{\phi}}{H} - 2 \frac{\dot{H}\dot{\phi}}{H^3} + 3 \frac{\ddot{\phi}}{H^2} \right) - 4 X (K_{\phi\phi XX} - G_{3\phi\phi\phi X}) \left( 2 \frac{\dot{H}\dot{\phi}}{H^3} - 3 \frac{\ddot{\phi}}{H^2} \right) \nonumber \\
& + 4 X G_{3\phi\phi X} \left( 9 - 9 \frac{\dot{H}}{H^2} - \frac{\dot{H}^2}{H^4} + 18 \frac{\ddot{\phi}}{H\dot{\phi}} + 4 \frac{\dot{H}\ddot{\phi}}{H^3\dot{\phi}} - 3 \frac{\ddot{\phi}^2}{H^2\dot{\phi}^2} \right) + 12 X G_{3\phi\phi\phi X} \frac{\dot{\phi}}{H} \nonumber \\
& + 8 X^2 G_{3\phi XXX} \left( 3 \frac{\dot{H}^2\dot{\phi}}{H^3} - 12 \frac{\dot{H}\ddot{\phi}}{H^2} - \frac{\dot{H}^2\ddot{\phi}}{H^4} + 9 \frac{\ddot{\phi}^2}{H\dot{\phi}} + 2 \frac{\dot{H}\ddot{\phi}^2}{H^3\dot{\phi}} - \frac{\ddot{\phi}^3}{H^2\dot{\phi}^2} \right) \nonumber \\
& + 8 G_{3\phi X} \Bigg( -9 \frac{\dot{H}\dot{\phi}}{H} + 3 \frac{\dot{H}^2\dot{\phi}}{H^3} + 4 \frac{\dot{H}^3\dot{\phi}}{H^5} - 3 \frac{\ddot{H}\dot{\phi}}{H^2} - 2 \frac{\dot{H}\ddot{H}\dot{\phi}}{H^4} + 9 \ddot{\phi} - 6 \frac{\dot{H}\ddot{\phi}}{H^2} \nonumber \\
&\qquad\qquad\quad - 9 \frac{\dot{H}^2\ddot{\phi}}{H^4} + 2 \frac{\ddot{H}\ddot{\phi}}{H^3} + 3 \frac{\ddot{\phi}^2}{H\dot{\phi}} + 6 \frac{\dot{H}\ddot{\phi}^2}{H^3\dot{\phi}} - \frac{\ddot{\phi}^3}{H^2\dot{\phi}^2} + 3 \frac{\dddot{\phi}}{H} + 2 \frac{\dot{H}\dddot{\phi}}{H^3} - 2 \frac{\ddot{\phi}\dddot{\phi}}{H^2\dot{\phi}} \Bigg) \nonumber \\
& + 4 X G_{3\phi XX} \Bigg( -18 \frac{\dot{H}\dot{\phi}}{H} + 21 \frac{\dot{H}^2\dot{\phi}}{H^3} + 4 \frac{\dot{H}^3\dot{\phi}}{H^5} - 6 \frac{\ddot{H}\dot{\phi}}{H^2} - 2 \frac{\dot{H}\ddot{H}\dot{\phi}}{H^4} + 18 \ddot{\phi} - 66 \frac{\dot{H}\ddot{\phi}}{H^2} \nonumber \\
&\qquad\qquad\qquad - 15 \frac{\dot{H}^2\ddot{\phi}}{H^4} + 45 \frac{\ddot{\phi}^2}{H\dot{\phi}} + 18 \frac{\dot{H}\ddot{\phi}^2}{H^3\dot{\phi}} - 7 \frac{\ddot{\phi}^3}{H^2\dot{\phi}^2} + 6 \frac{\dddot{\phi}}{H} + 2 \frac{\dot{H}\dddot{\phi}}{H^3} - 2 \frac{\ddot{\phi}\dddot{\phi}}{H^2\dot{\phi}} + 2 \frac{\ddot{H}\ddot{\phi}}{H^3} \Bigg) \Bigg].
\end{align}
\end{widetext}

\bibliographystyle{utphys}
\bibliography{MasterScreening}

\end{document}